\documentclass[a4paper, 11pt]{article}
\usepackage{epsf}
\usepackage{epsfig}
\usepackage{psfrag}
\usepackage{amsmath}
\usepackage{amssymb}
\usepackage{theorem}
\usepackage{natbib}
\begin{document}
\title{A Machine Learning Strategy to Identity Exonic Splice Enhancers in
  Human Protein-coding Sequence}
\author{Thomas A. Down, Bernard Leong and Tim J. P. Hubbard \\ 
        Wellcome Trust Sanger Institute, Hinxton, Cambridge CB10 1SA}
\date{\today}
\maketitle

\begin{abstract}
{\bf Background:} Exonic splice enhancers are sequences embedded within exons
which promote and regulate the splicing of the transcript in which they are
located. A class of exonic splice enhancers are the SR proteins, which
are thought to mediate interactions between splicing factors bound to
the 5' and 3' splice sites. 

{\bf Method and results:} We present a novel strategy for analysing protein-coding
sequence by first randomizing the codons used at each position within the coding
sequence, then applying a motif-based machine learning algorithm to compare the
true and randomized sequences.  This strategy identified a collection of motifs
which can successfully discriminate between real and randomized coding sequence,
including -- but not restricted to -- several previously reported splice enhancer
elements.  As well as successfully distinguishing coding exons from randomized sequences, 
we show that our model is able to recognize non-coding exons.

{\bf Conclusions:} Our strategy succeeded in detecting signals in coding exons
which seem to be orthogonal to the sequences' primary function of coding for
proteins.  We believe that many of the motifs detected here may represent
binding sites for previously unrecognized proteins which influence RNA splicing.  We hope that this
development will lead to improved knowledge of exonic splice enhancers, and
new developments in the field of computational gene prediction.

\end{abstract}

\section{Introduction} \label{intro}
Alternative splicing is a major mechanism of diversity in the expression
of eukaryotic genes, and has also been implicated in gene regulation
\citep{modrek.lee, brett.etal, graveley2001, harrison.etal, lewis.green.brenner}.  A number of sequences have been found embedded 
in the exons of both viral and cellular
genes which can promote or repress the utilization of alternative splice
sites.  Exonic splice enhancers are usually purine-rich
sequences located in an alternatively splice donor.  Through specific binding of proteins, 
including the serine/arginine-rich SR family, exonic splice enhancers function
by recruiting splicing factors such as U2AF to a suboptimal splice acceptor
in the early stages of splicesome assembly, thereby stimulating
splicing of the upstream intron or inclusion of the alternative exon.

SR proteins are a family of highly conserved serine/arginine-rich
RNA-binding proteins (For review, see \citealt{graveley}). They are
essential splicing factors and also regulate the selection and use of
alternative splice sites \citep{bourgeois.etal, liu.zhang.krainer, 
lynch.etal, schaal.maniatis, tacke.manley, tacke.tohyama.ogawa, tian.kole, 
zheng.huynen.baker}. It is known that these proteins function very early in the spliceosome
assembly process. They promote the binding of U1 snRNP to the splice donor
and of U2AF to the splice acceptor, apparently by interacting with U1 70K
and U2AF respectively. Observations have shown that SR proteins
bound to the exonic splice enhancers recruit splicing factors
to the adjacent splice sites. There are nine human SR
proteins which are presently known and studied: SF2/ASF \citep{graveley, liu.zhang.krainer, schaal.maniatis, tacke.manley}, 
SC35 \citep{liu.zhang.krainer, tacke.manley, schaal.maniatis, tacke.tohyama.ogawa}, 
SRp20 \citep{schaal.maniatis, tacke.tohyama.ogawa}, SRp40 \citep{liu.zhang.krainer}, 
SRp55 \citep{liu.zhang.krainer}, SRp75\citep{graveley}, SRp30c \citep{graveley}, 9G8
\citep{schaal.maniatis, tian.kole} and the divergent
SRp54 \citep{graveley}.  These proteins are closely 
related in sequence and structure and share the ability to activate
splicing. Another class of human SR related proteins, the Tra2 family,
are also known to be splicing regulators
and sequence specific activators of pre-mRNA splicing \citep{tacke.tohyama.ogawa}. 

Early research concentrated on how SR proteins function to regulate
alternative splicing.  However, the binding of SR proteins to
constitutive exons -- those which are included in all splice variants of a gene --
also plays an important role in the splicing
reaction. The exon definition model proposes that interactions between components bound to 
splice sites flanking an exon serve to highlight exons -- which are
usually small -- against a background of much larger introns. It is conjectured that the
majority of constitutively spliced exons are defined by this
mechanism. To support the model, a number of SR protein binding sites
have been identified in constitutive exons, and also shown to be
constitutive splicing enhancers \citep{schaal.maniatis, lam.hertel}.

Although examples of exonic splice enhancers are believed to be common, studying
their sequences is difficult because they are embedded in exons, most of which
are also functional protein-coding sequences. Non-coding exons are also thought
to contain many sequences which are functional in the mature RNA, such as
regulators of RNA stability, so the situation there is not necessarily clearer.
When a particular motif is found to be over- or under-represented in coding
exons, it is generally unclear whether it is a consequence of the underlying
protein sequence, or an unrelated signal -- such as a splice enhancer --
embedded in the protein coding sequence. Here we propose a novel strategy for
resolving this uncertainty. Starting with annotated coding exons, we generate a
`neutralized' exon set: sequences which are generated randomly, but which
nevertheless preserve both the amino acid sequence and overall composition
features of the true exons. We then apply machine learning software to compare
the true and neutralized exons. Since the neutralized set codes for the same
proteins, it is likely that any feature which can be used to discriminate
between the true and neutralized sets is performing some function which is
independent of the exons' primary, protein-coding, function.

The neutralization process we use has some similarities to the
dicodon shuffling algorithm proposed by \citet{katz.burge}, which
swaps pairs of synonymous codons under a constraint that the dinucleotide
composition of the sequence must be preserved.  However, our method differs
both in implementation strategy and in the fact that dinucleotide composition
is maintained across the complete set of sequences, rather than on a per-sequence
basis (see results in section \ref{neutralizedexons}).

An alternative, very different, computational method for finding splice enhancer
signals has recently been proposed: RESCUE-ESE \citep{fairbrother.etal} compares
the sequences around weak consensus splice sites with those around strong consensus
sites, with the expectation that splice enhancer motifs are more likely to be found
in the vicinity of weak splice sites.  This strategy is very different from ours,
and so it is interesting to compare the results.

\section{Results}

\subsection{Neutralized exons} \label{neutralizedexons}

\begin{figure}[!bth]
\begin{center}
\includegraphics[scale=1.0]{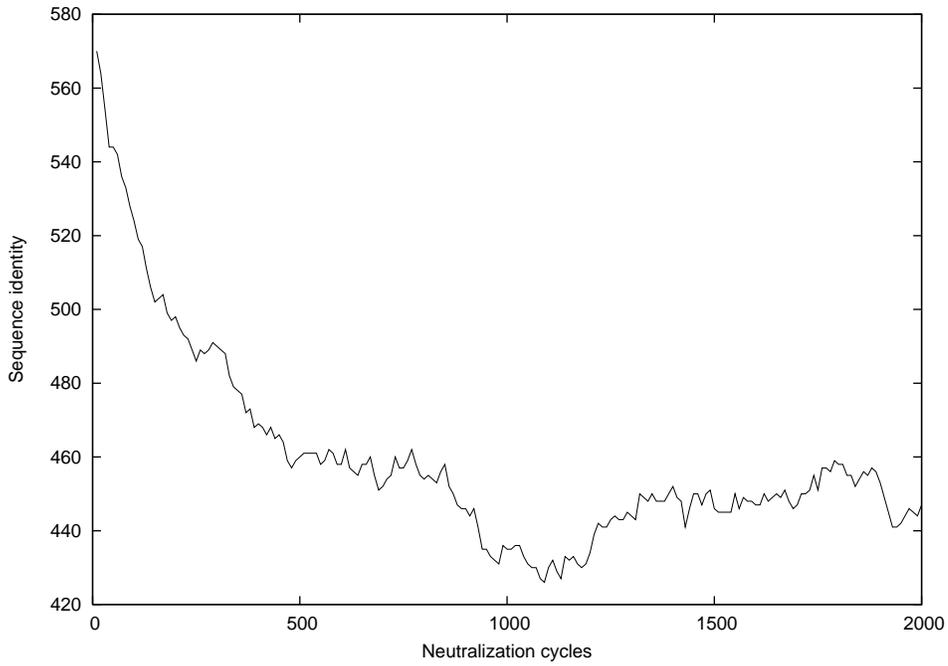}
\caption{Time-course for neutralizing a typical 300 base sequence.}
\label{seqid}
\end{center}
\end{figure}

Internal coding exons with lengths ranging from 100 to 300 bases were extracted
from the Vega database of annotated human genomic sequence [http://vega.sanger.ac.uk/].
Testing the neutralization process on a typical 300 base exon (figure \ref{seqid})
we see that the level of sequence identity falls
steadily for approximately 500 cycles, then comes close to its minimum
value and only fluctuates slightly for the remainder of the cycles.  Allowing some
margin for exceptional sequences, this suggests that
that 1000 cycles of neutralization is adequate to randomize any sequence with
a length up to 300 bases.

Running the neutralization algorithm on the complete set of qualifying exons, for
1000 cycles per exon, gave a neutralized set of with an average of 78\% identity.
The average dinucleotide compositions of the exons before and after neutralization
is shown in table \ref{tdinuc}.  We can see that most dinucleotides show negligible
change in composition during the neutralization procedure, and in the most extreme
case (the tt dinucleotide), the proportion of the sequences composed of tt
dinucleotides changes by less that 2\%.  Therefore, we consider the neutralization
algorithm to be successful in preserving overall sequence composition while substantially
changing the sequence itself. On the same sequences, the dicodon shuffling algorithm
typically gives a sequence identity of around 90\%.

\begin{table}[!bth]
\begin{center}
\begin{tabular}{| p{3cm} | p{3cm} | p{3cm} |} 
\hline
Dinucleotide    &   True exons  & Neutralized exons\\
\hline
aa  &   7.71\%  &   7.73\%\\
ac  &   5.56\%  &   5.55\%\\
ag  &   8.19\%  &   8.21\%\\
at  &   5.58\%  &   5.62\%\\
ca  &   8.06\%  &   8.06\%\\
cc  &   7.17\%  &   7.18\%\\
cg  &   2.73\%  &   2.74\%\\
ct  &   6.96\%  &   6.88\%\\
ga  &   7.76\%  &   7.76\%\\
gc  &   6.41\%  &   6.38\%\\
gg  &   6.61\%  &   6.54\%\\
gt  &   4.55\%  &   4.58\%\\
ta  &   3.43\%  &   3.47\%\\
tc  &   5.72\%  &   5.69\%\\
tg  &   7.98\%  &   7.96\%\\
tt  &   5.48\%  &   5.58\%\\
\hline
\end{tabular} 
\caption{Comparison of dinucleotide frequencies in true and neutralized exons}
\label{tdinuc}
\end{center}
\end{table}

\subsection{Motif-based models can effectively distinguish between true and neutralized exons}

Our data set consisted of 9091 true coding exons ranging in length from 100 to 300
bps, and an equal number of neutralized counterparts (see methods section).  From both the true
and neutralized sets, we removed 300 randomly selected sequences for use
as test data.  The remainder were used to train a Convolved Eponine Windowed Sequence (C-EWS)
model (see methods and \citealt{down.rvmseq}).  These models are based on scaffolds
of one of more sequence motifs (in this case,
limited to a maximum of three per scaffold).  The motifs are represented as DNA
weight matrices \citep{bucher.wms}.  When a scaffold includes more than one motif, probability
distributions associated with each motif indicate the preferred relative positions
of those motifs.  Each scaffold has an associated weight, which is used to combine
scaffold scores in a generalized linear model.

\begin{figure}[!bth]
\begin{center}
\begin{tabular}{p{4cm} p{4cm} p{4cm}}
\includegraphics[scale=0.2]{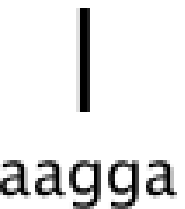}
\newline
\includegraphics[scale=0.2]{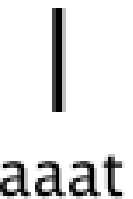}
\newline
\includegraphics[scale=0.2]{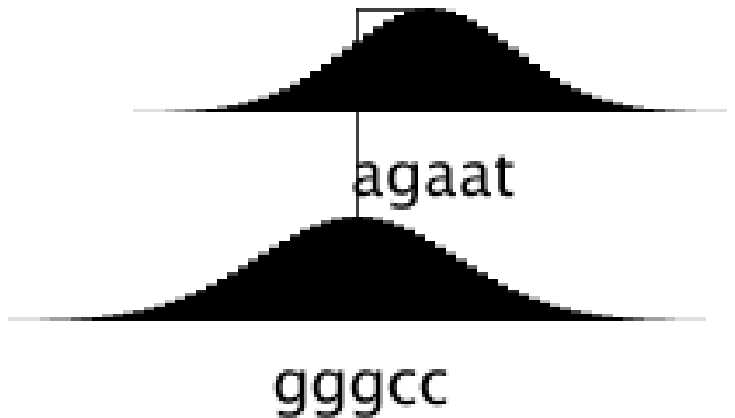}
\newline
\includegraphics[scale=0.2]{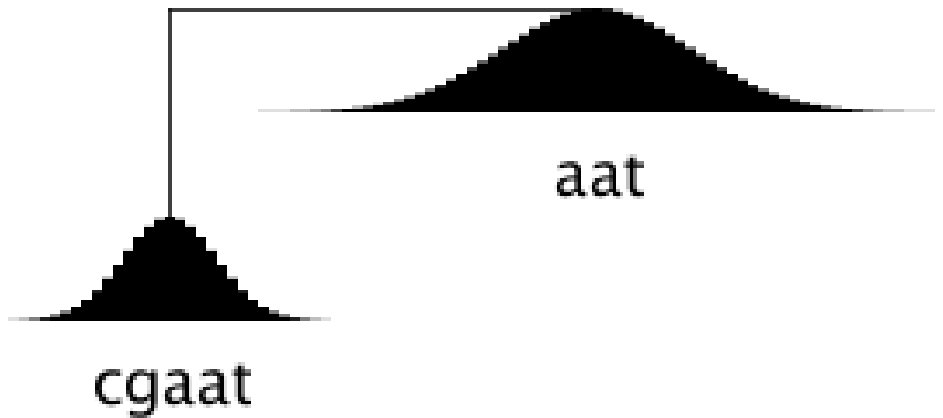}
\newline
\includegraphics[scale=0.2]{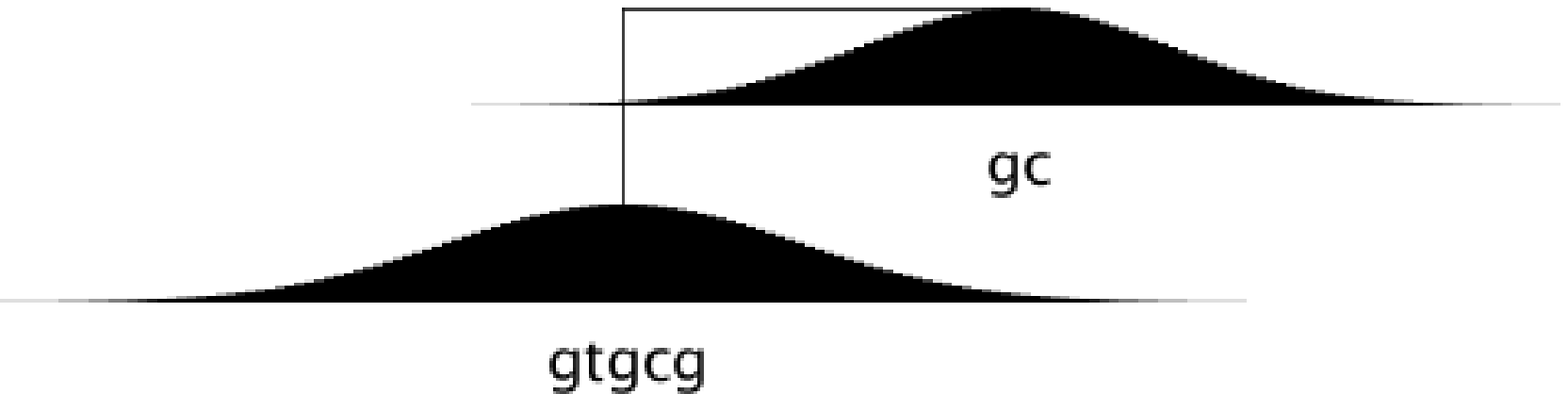}
\newline
\includegraphics[scale=0.2]{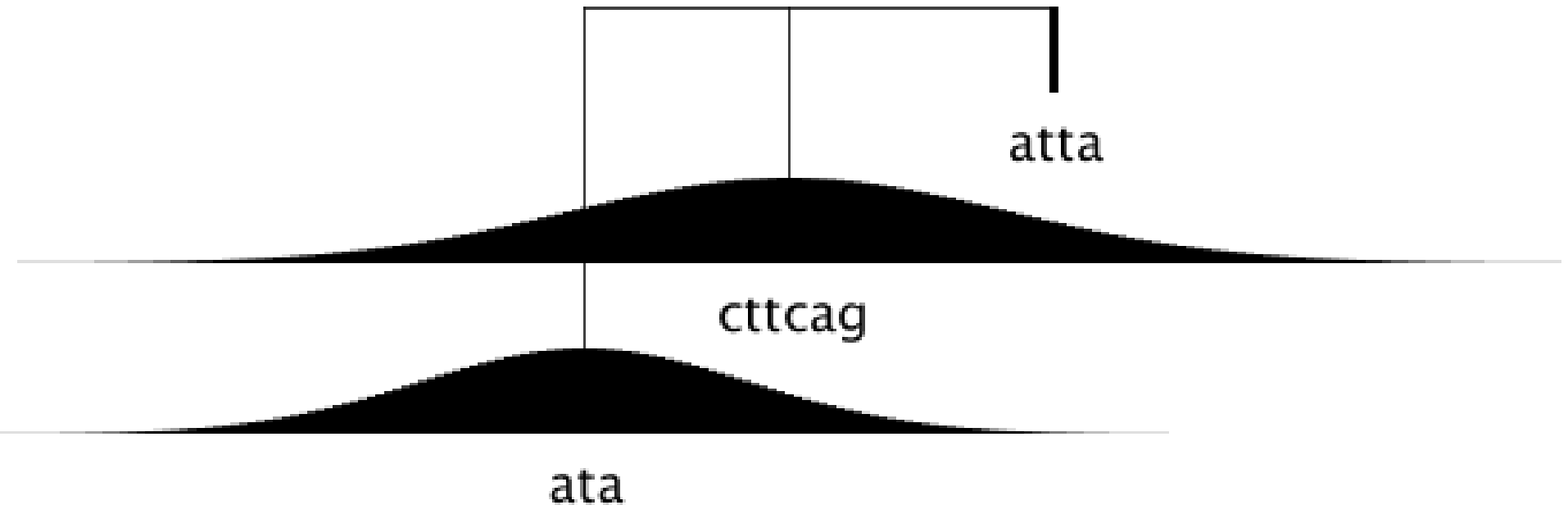}
\newline
\includegraphics[scale=0.2]{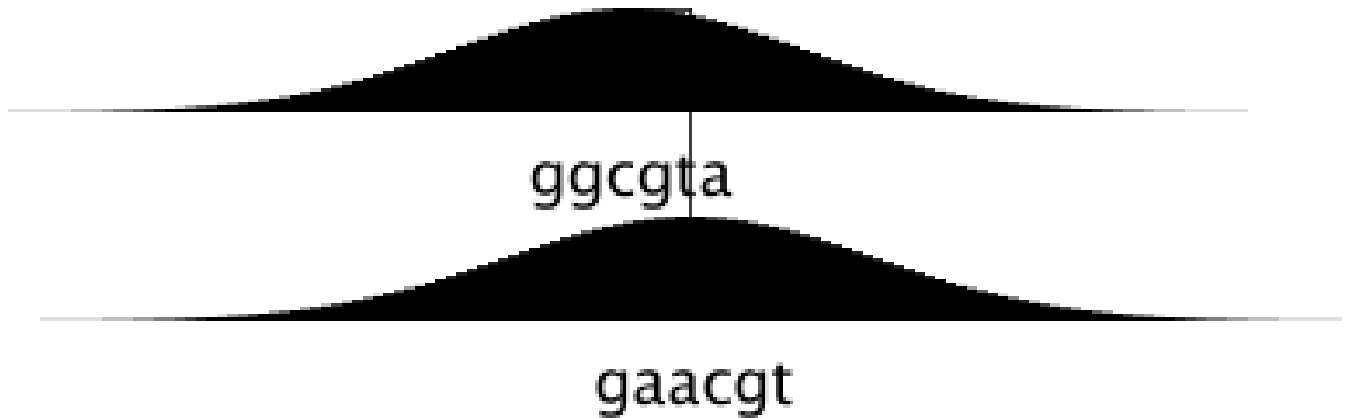}
\newline
\includegraphics[scale=0.2]{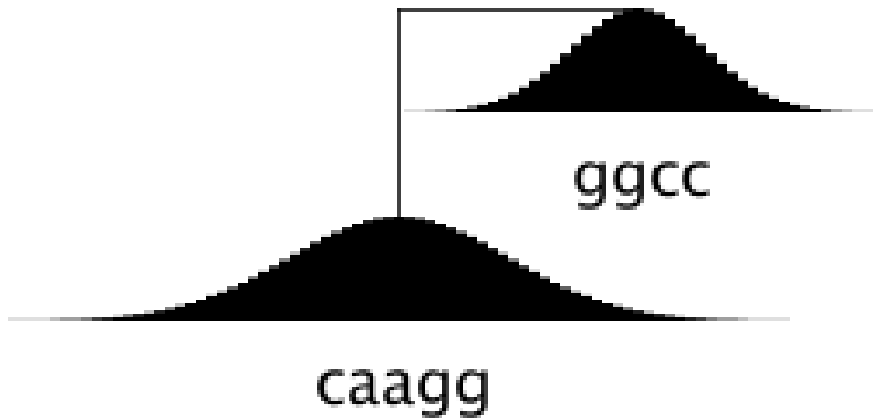}
\newline
\includegraphics[scale=0.2]{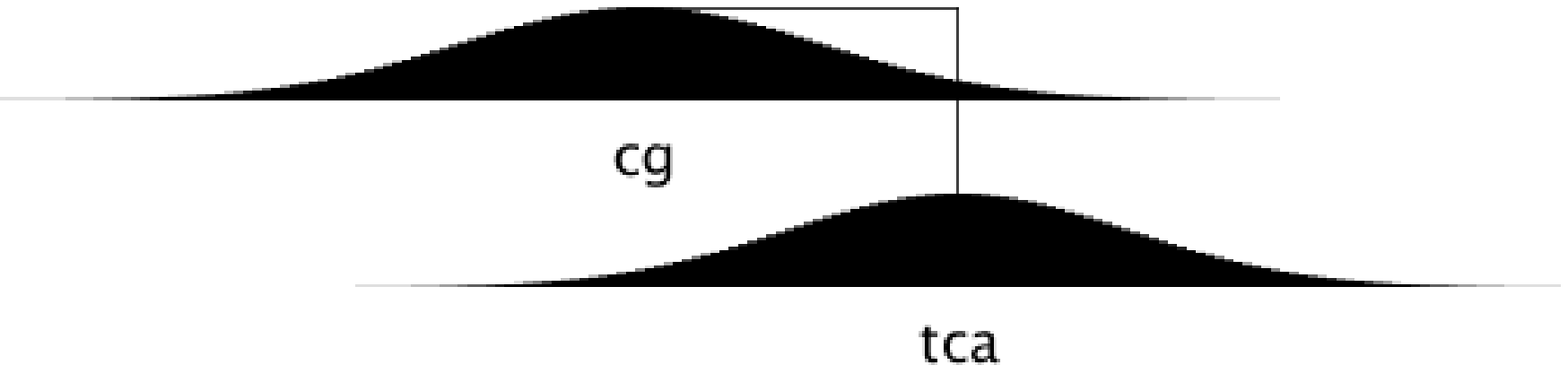}
\newline
\includegraphics[scale=0.2]{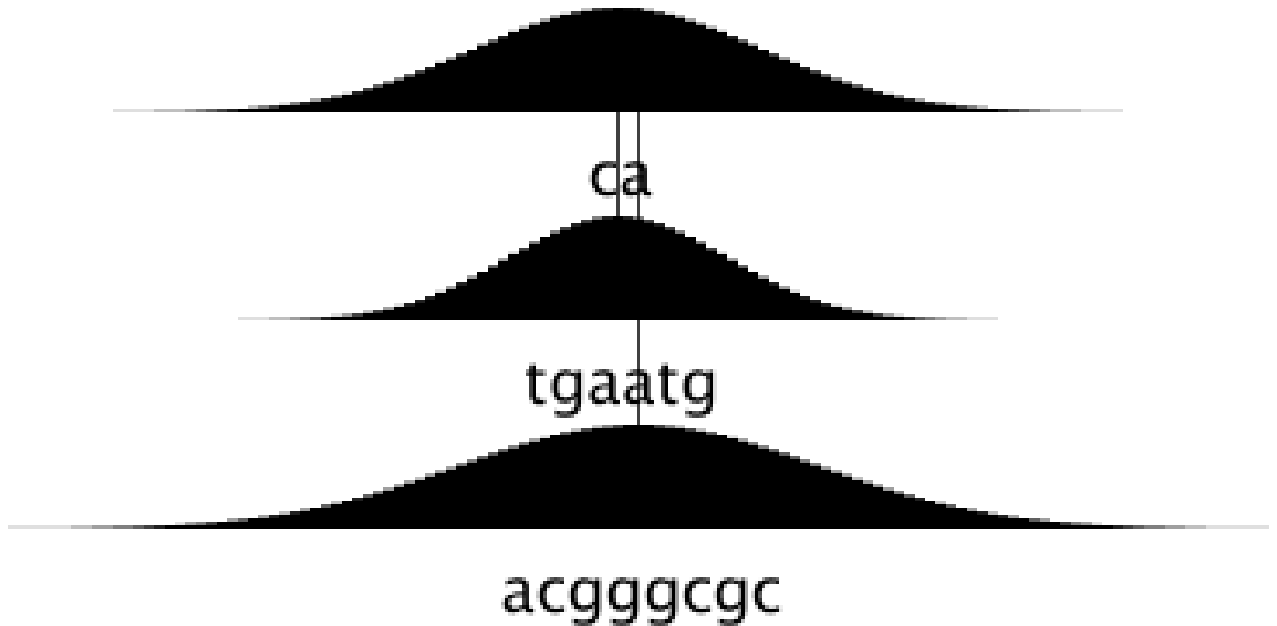}
&
\includegraphics[scale=0.2]{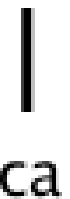}
\newline
\includegraphics[scale=0.2]{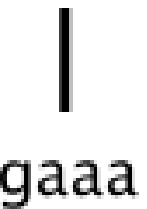}
\newline
\includegraphics[scale=0.2]{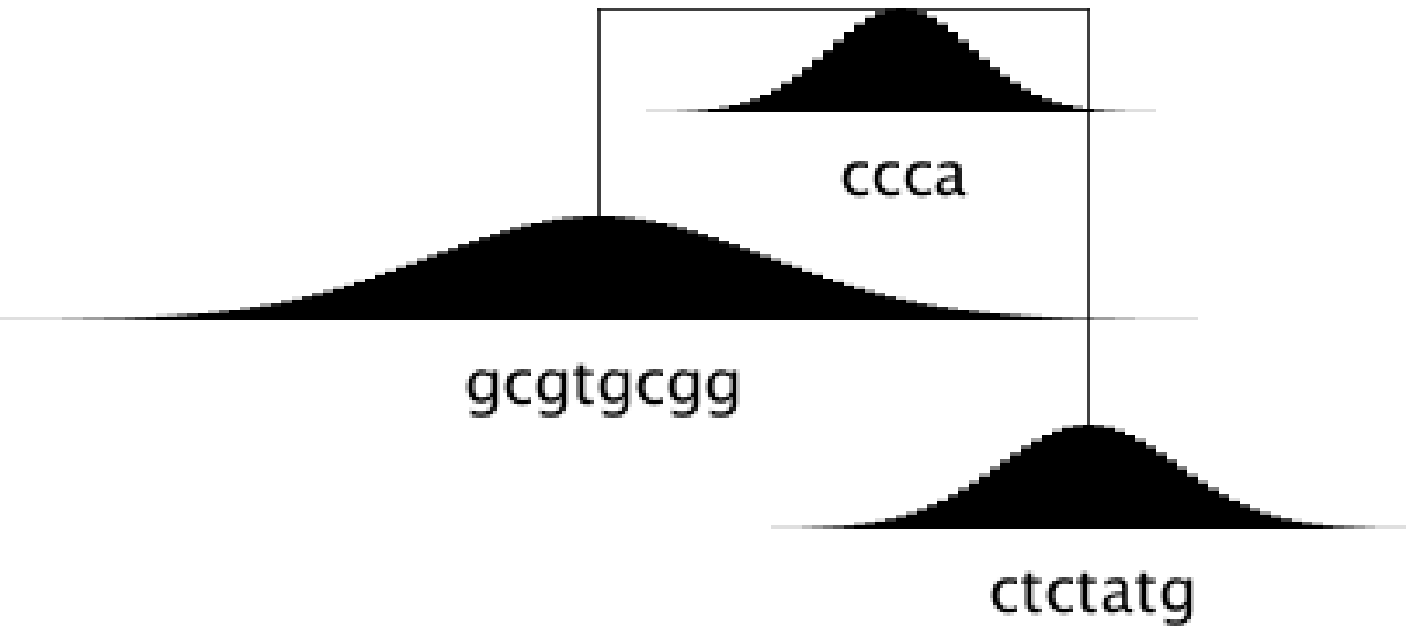}
\newline
\includegraphics[scale=0.2]{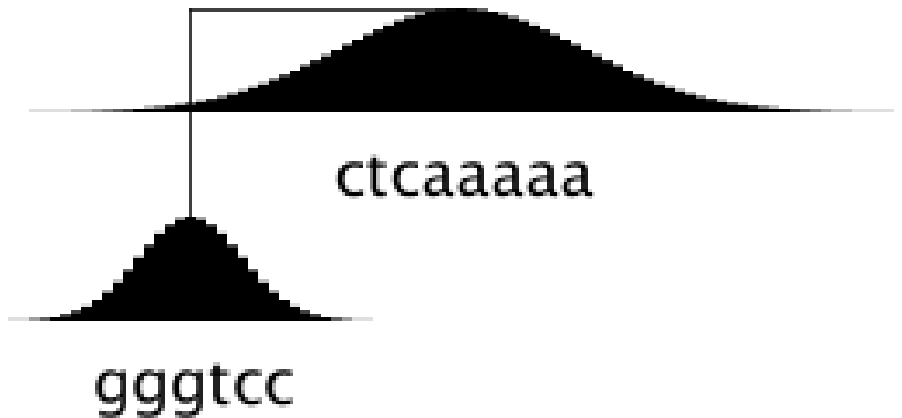}
\newline
\includegraphics[scale=0.2]{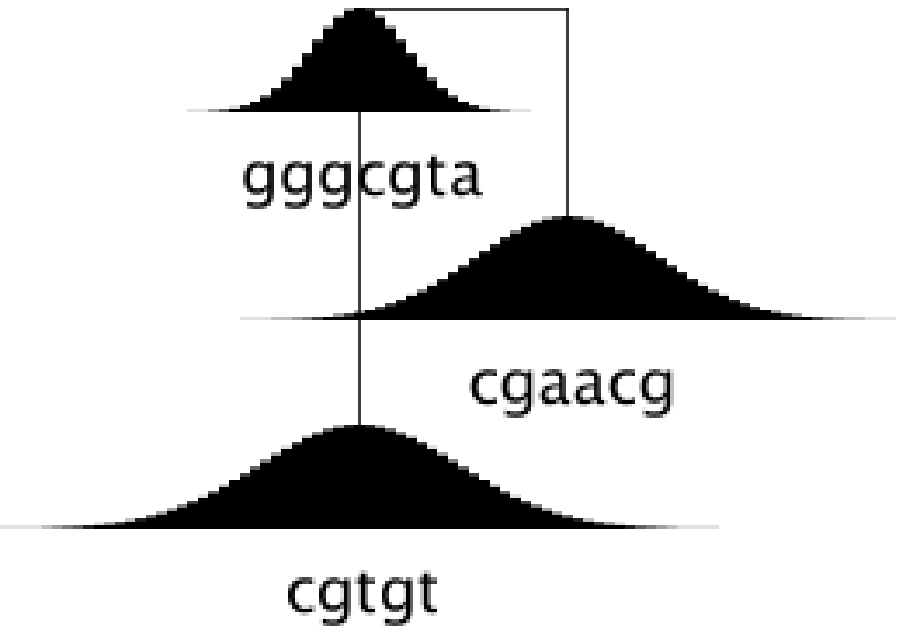}
\newline
\includegraphics[scale=0.2]{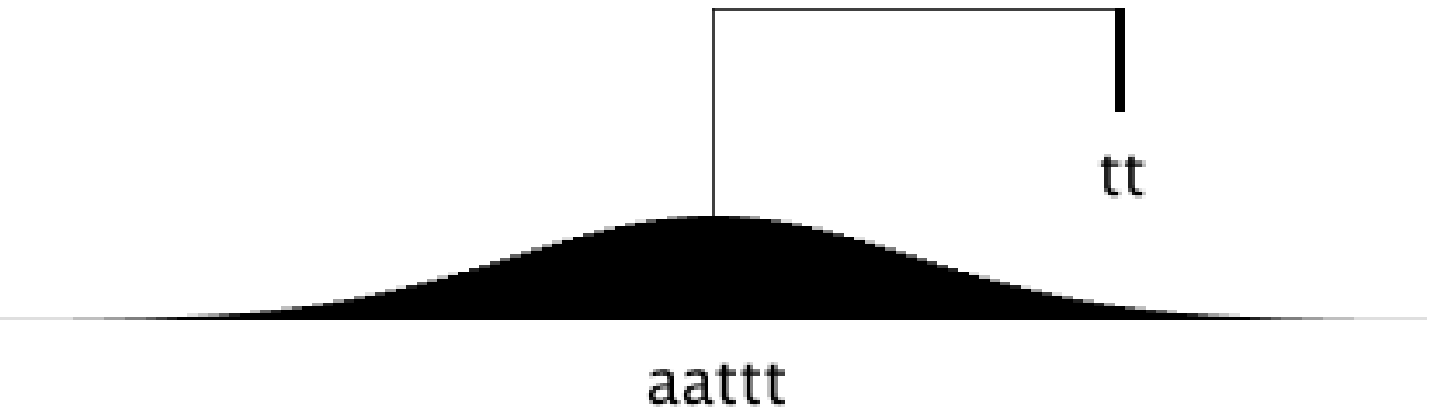}
\newline
\includegraphics[scale=0.2]{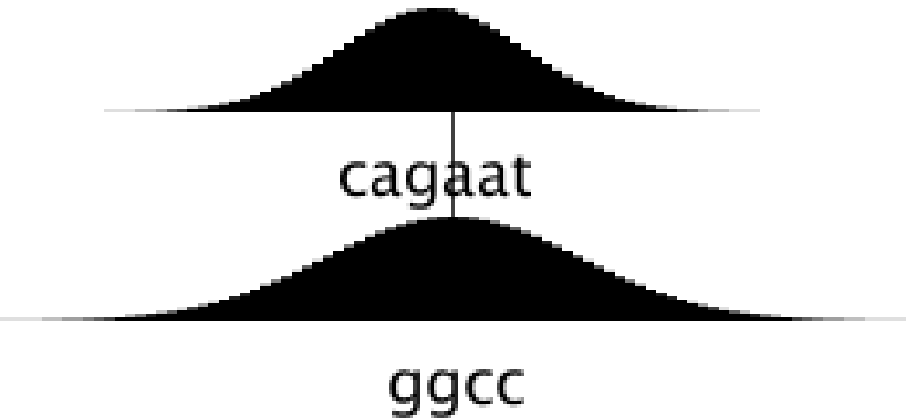}
\newline
\includegraphics[scale=0.2]{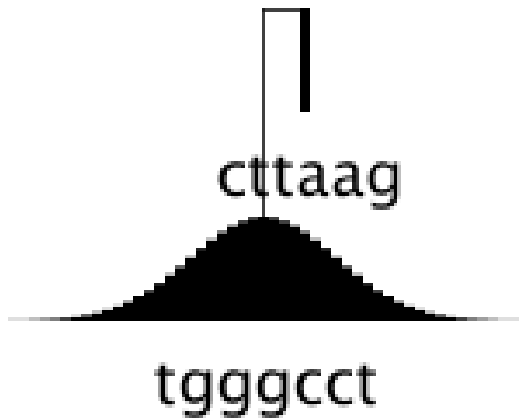}
\newline
\includegraphics[scale=0.2]{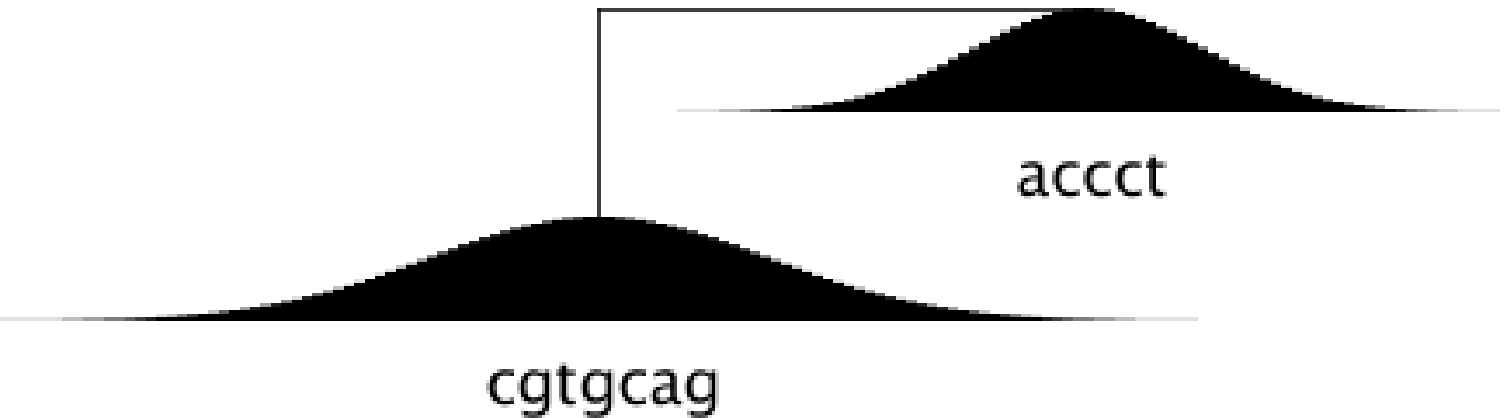}
\newline
\includegraphics[scale=0.2]{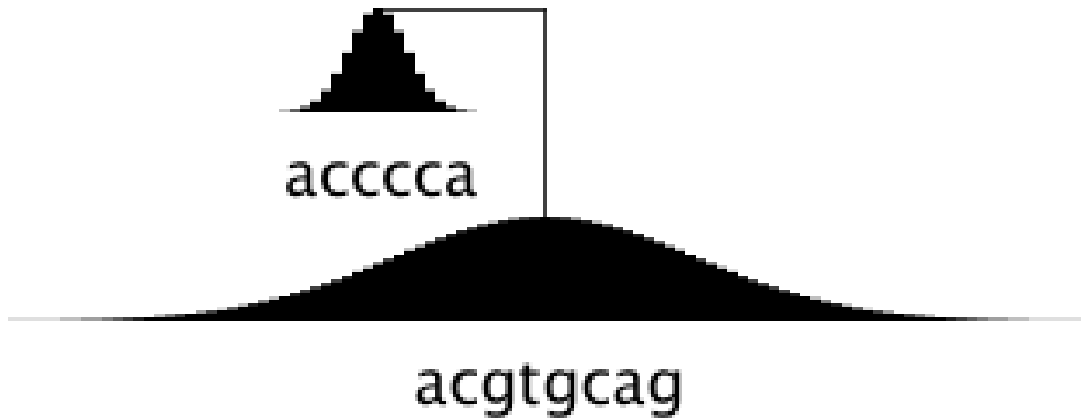}
&
\includegraphics[scale=0.2]{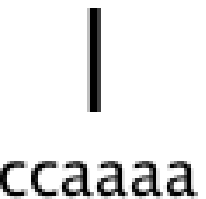}
\newline
\includegraphics[scale=0.2]{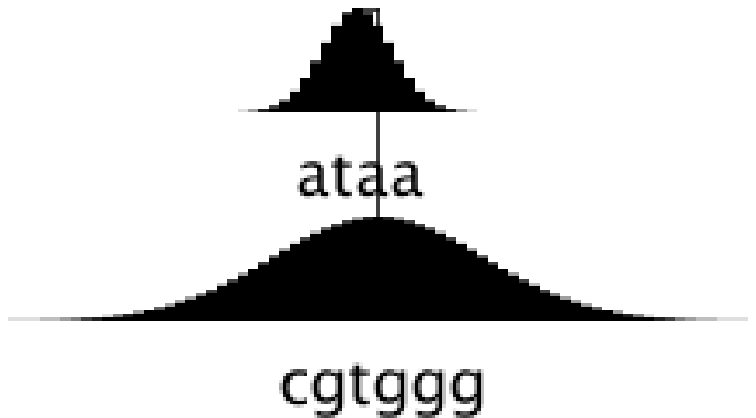}
\newline
\includegraphics[scale=0.2]{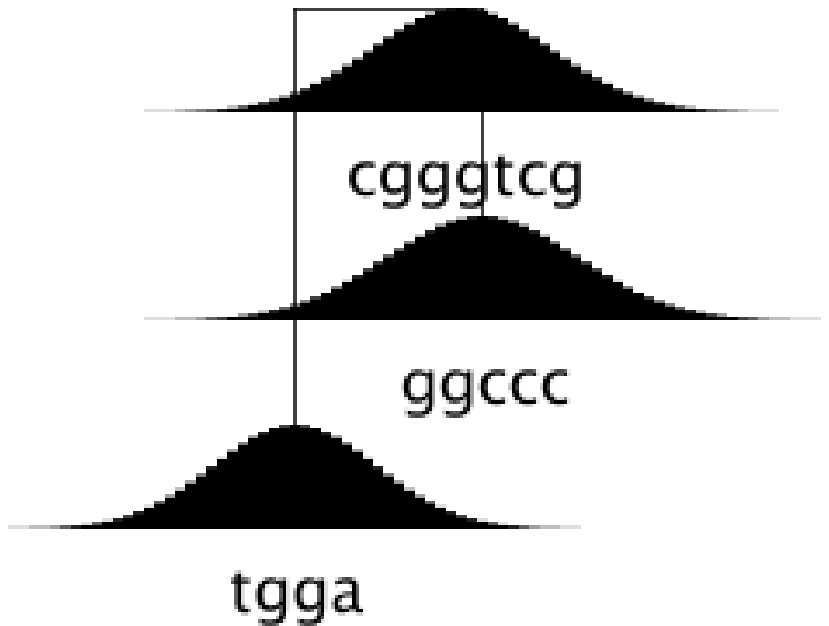}
\newline
\includegraphics[scale=0.2]{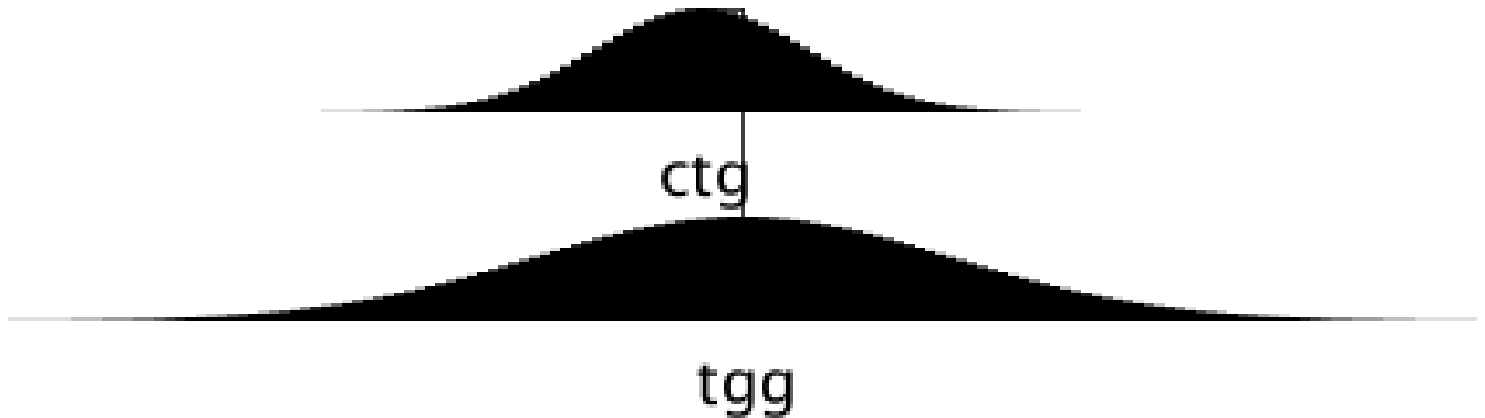}
\newline
\includegraphics[scale=0.2]{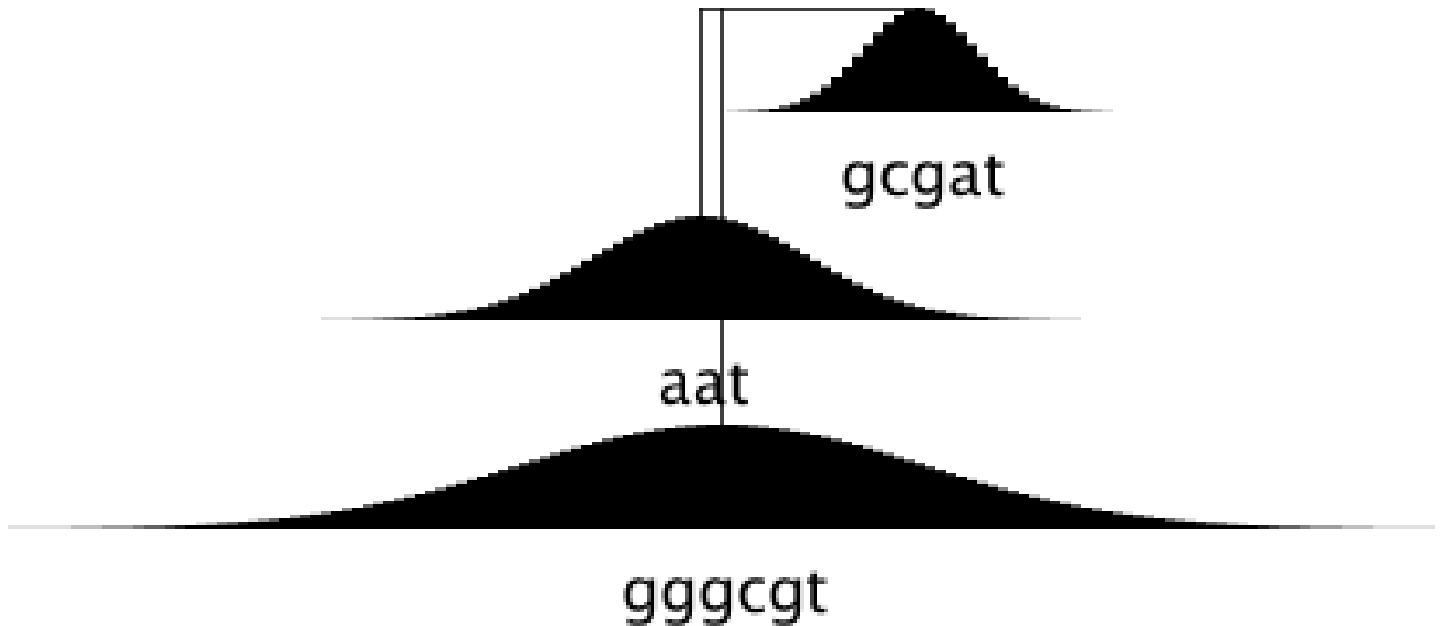}
\newline
\includegraphics[scale=0.2]{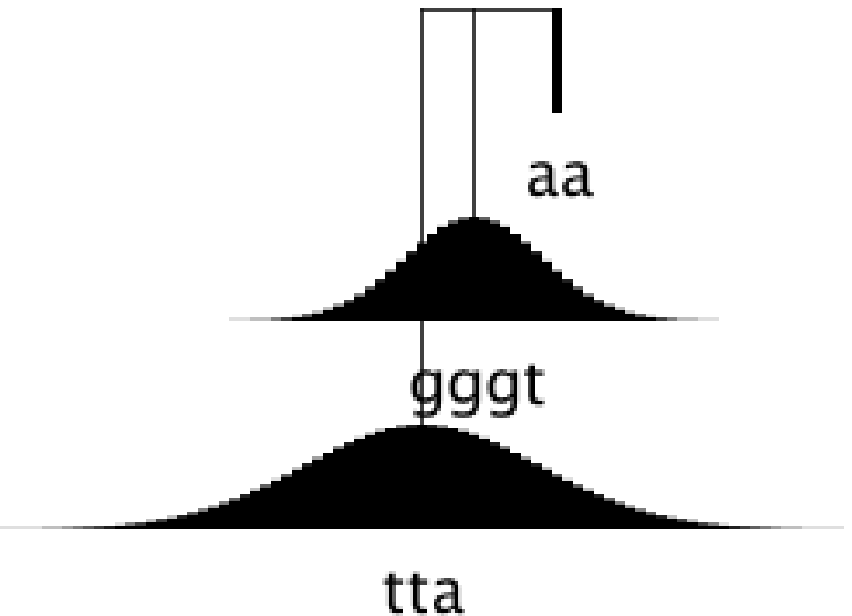}
\newline
\includegraphics[scale=0.2]{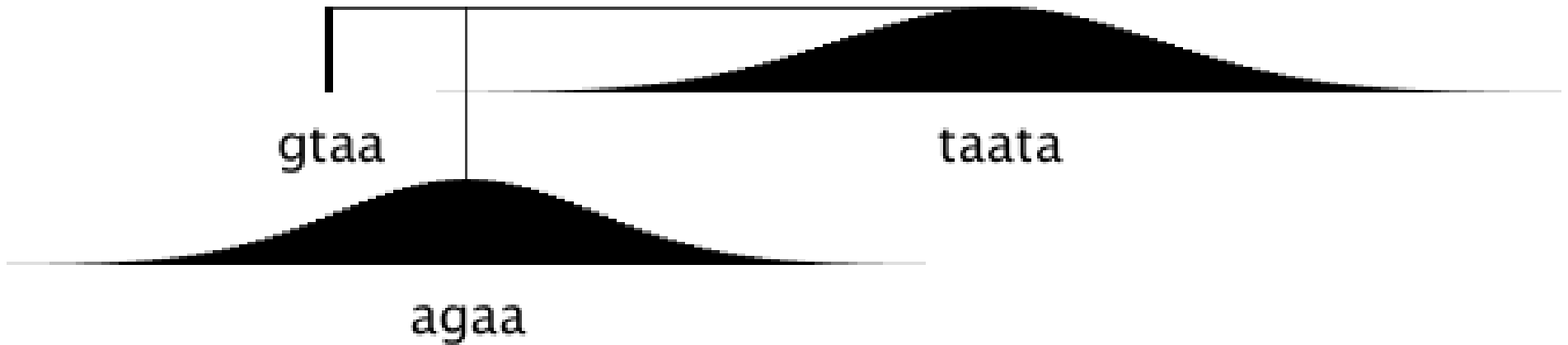}
\newline
\includegraphics[scale=0.2]{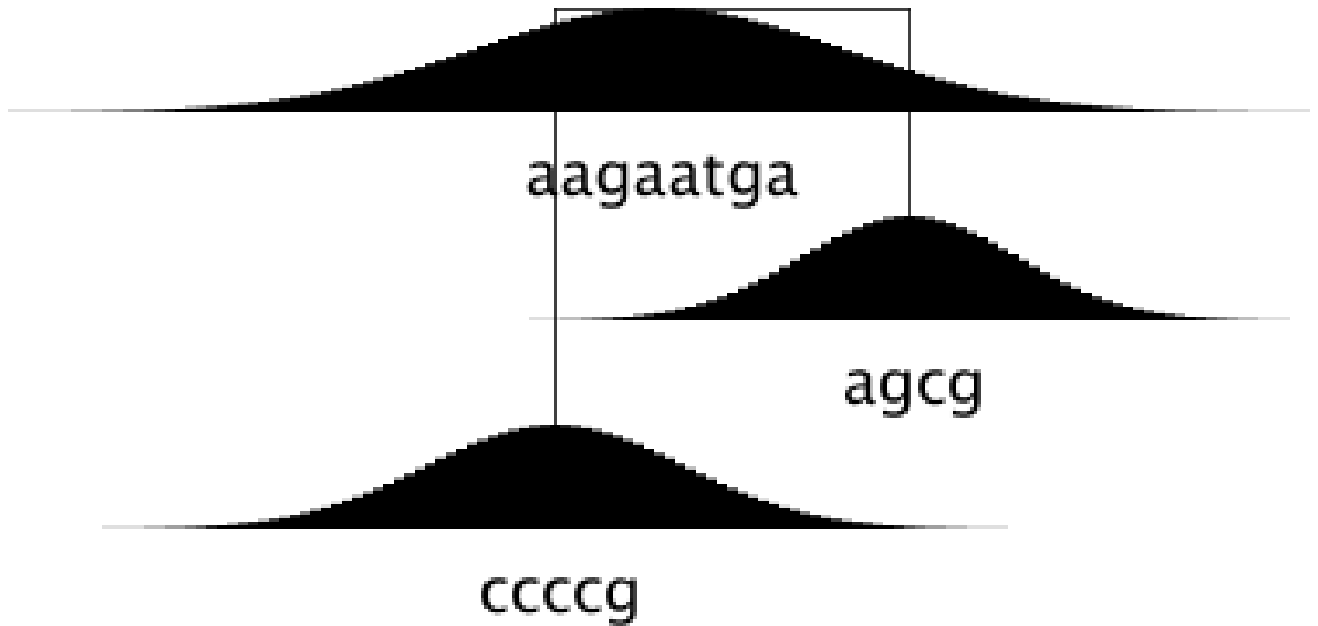}
\newline
\includegraphics[scale=0.2]{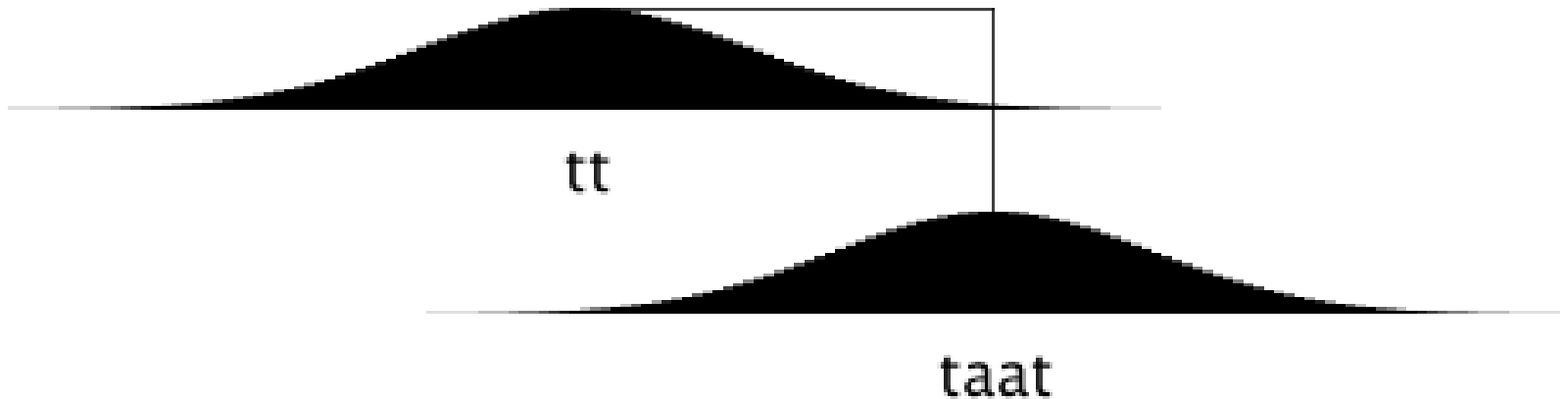}
\newline
\includegraphics[scale=0.2]{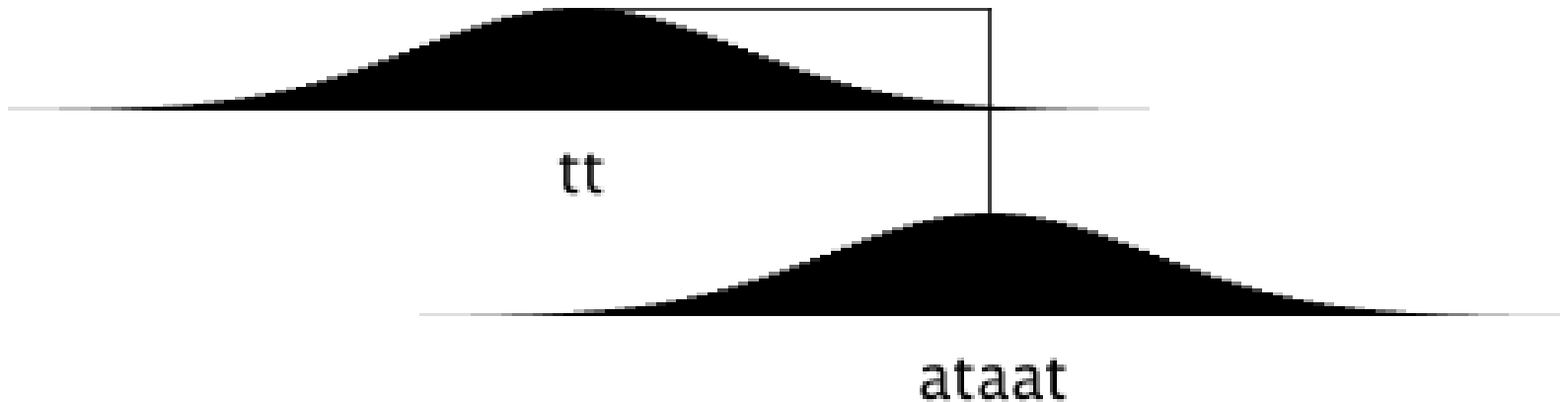}
\end{tabular}

\caption{Positively weighted scaffolds in the Eponine Exons model}
\label{positive.scaffolds}
\end{center}
\end{figure}

\begin{figure}[!bth]
\begin{center}
\begin{tabular}{p{4cm} p{4cm} p{4cm}}
\includegraphics[scale=0.2]{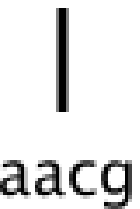}
\newline
\includegraphics[scale=0.2]{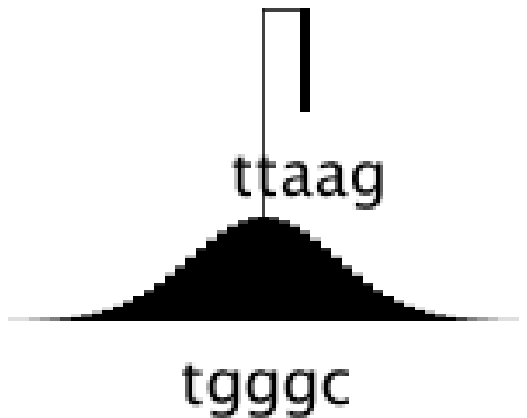}
\newline
\includegraphics[scale=0.2]{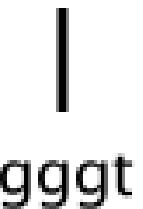}
\newline
\includegraphics[scale=0.2]{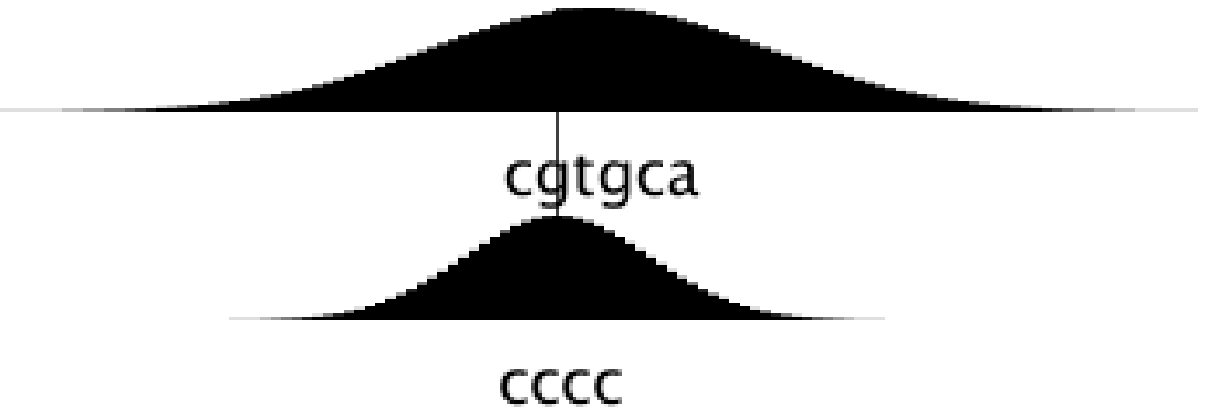}
\newline
\includegraphics[scale=0.2]{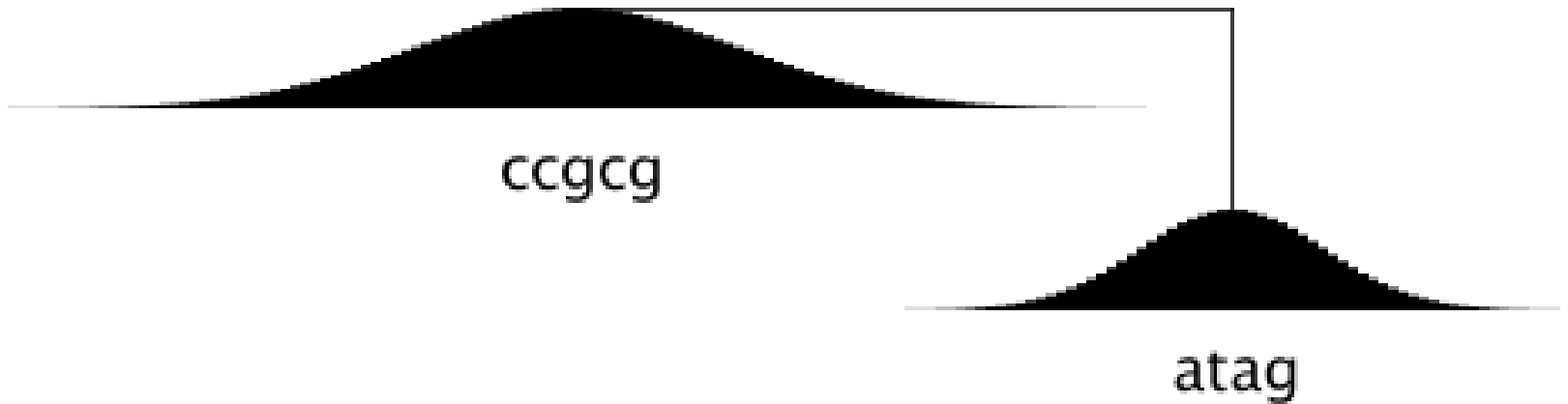}
\newline
\includegraphics[scale=0.2]{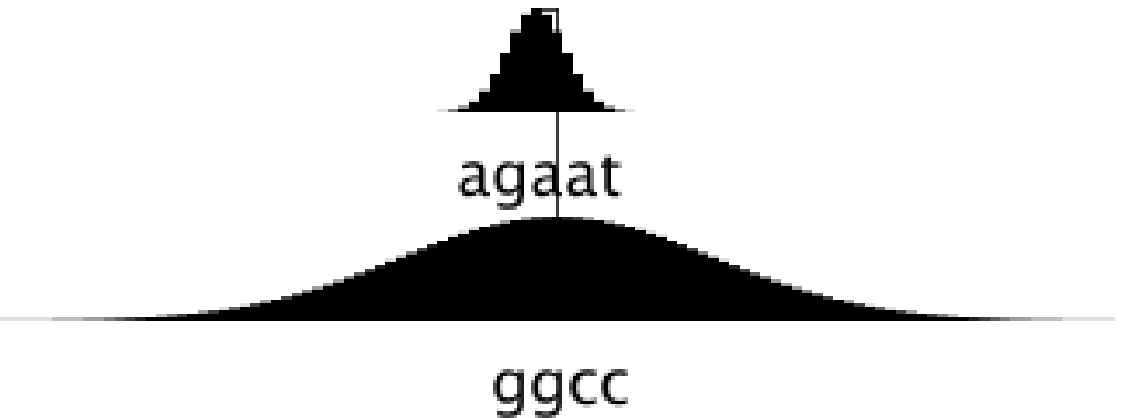}
\newline
\includegraphics[scale=0.2]{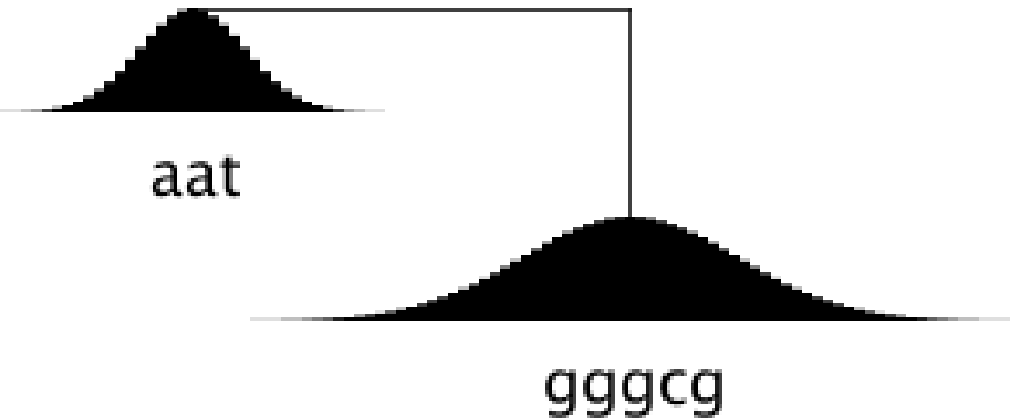}
\newline
\includegraphics[scale=0.2]{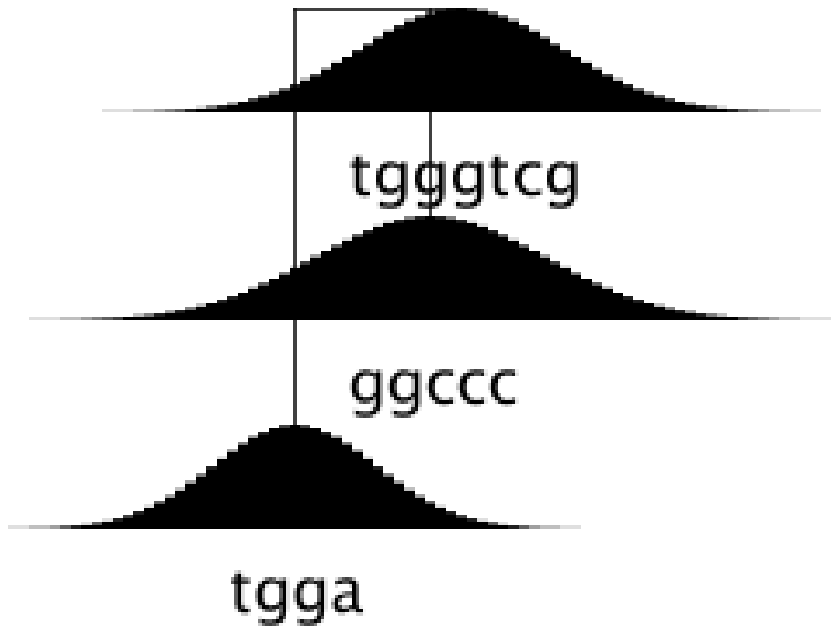}
\newline
\includegraphics[scale=0.2]{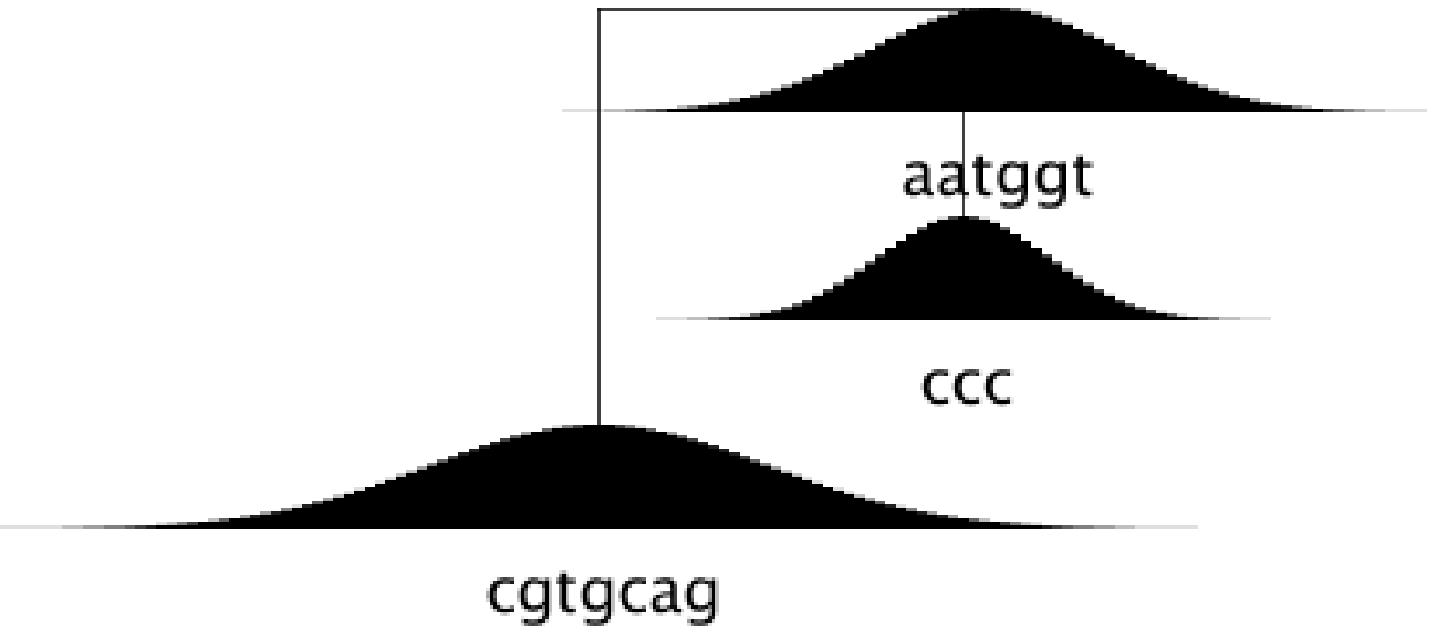}
\newline
\includegraphics[scale=0.2]{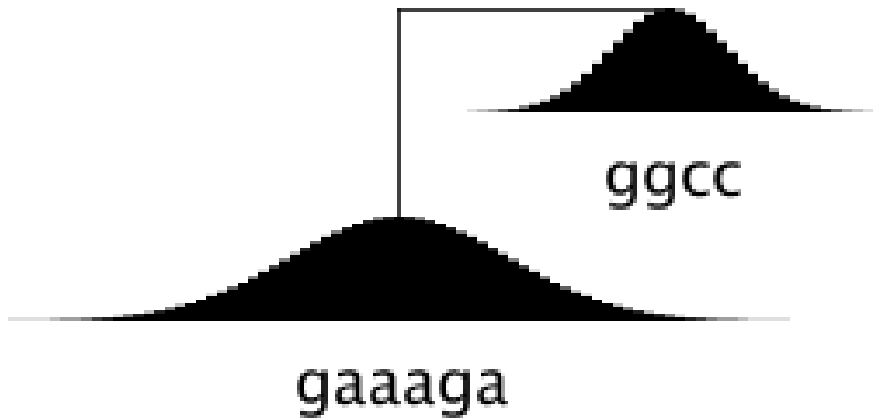}
\newline
\includegraphics[scale=0.2]{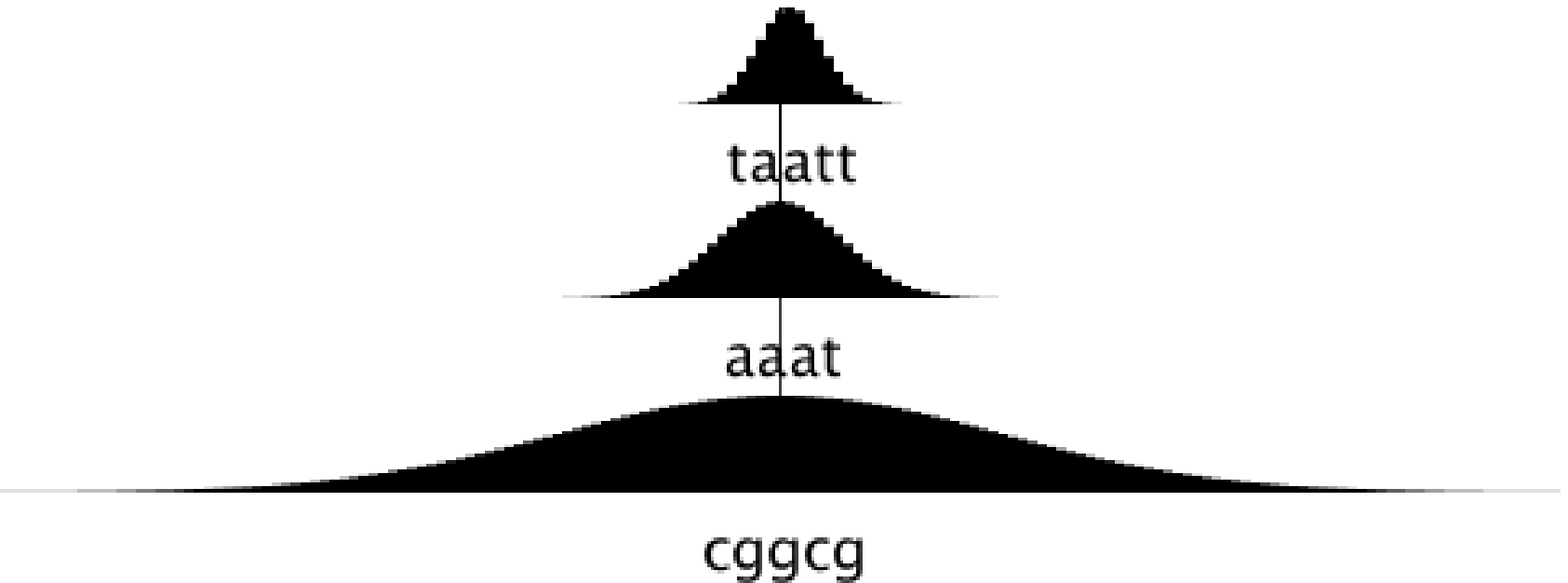}
\newline
\includegraphics[scale=0.2]{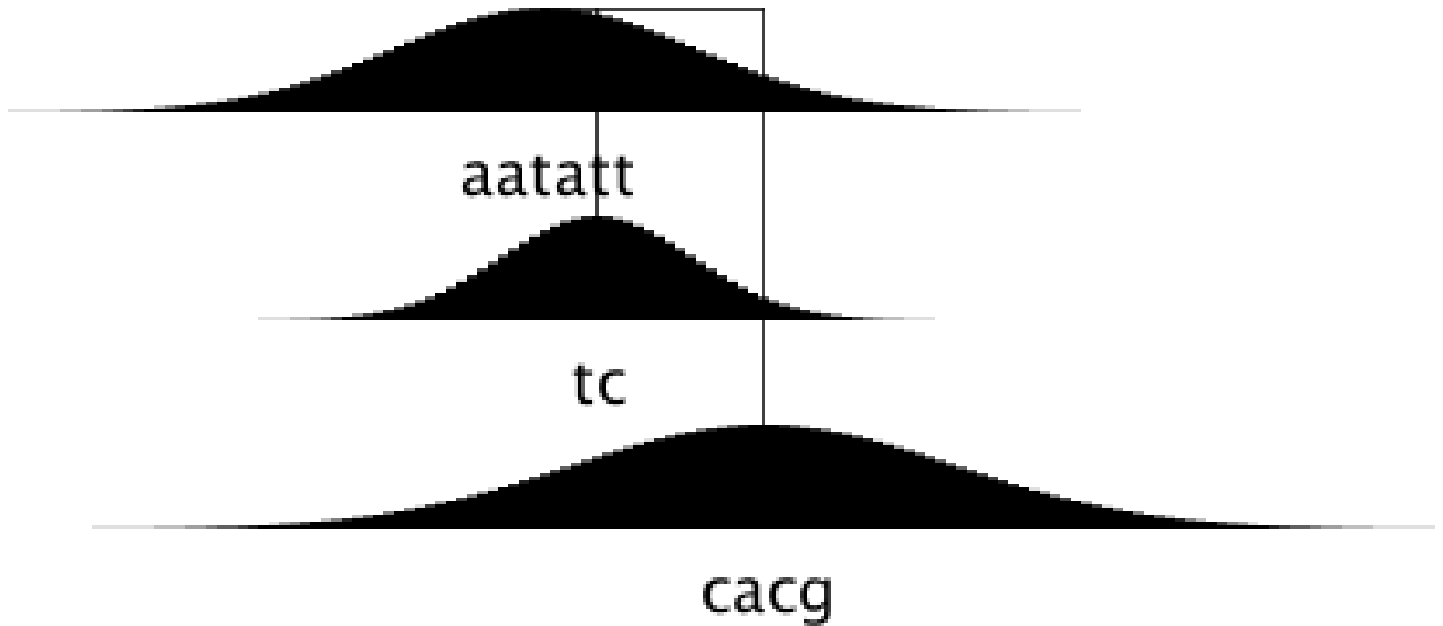}
\newline
\includegraphics[scale=0.2]{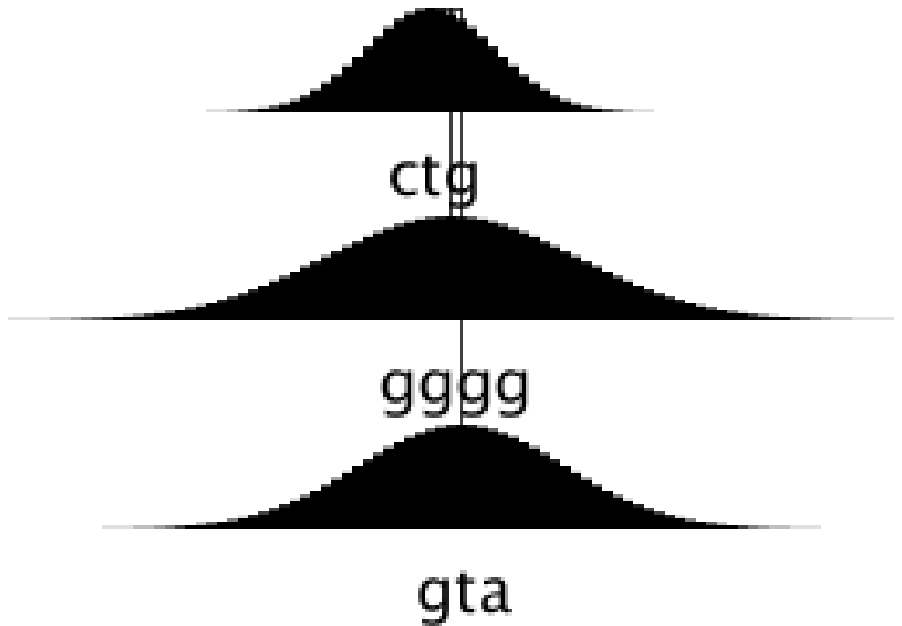}
&
\includegraphics[scale=0.2]{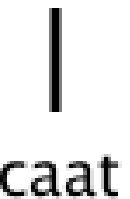}
\newline
\includegraphics[scale=0.2]{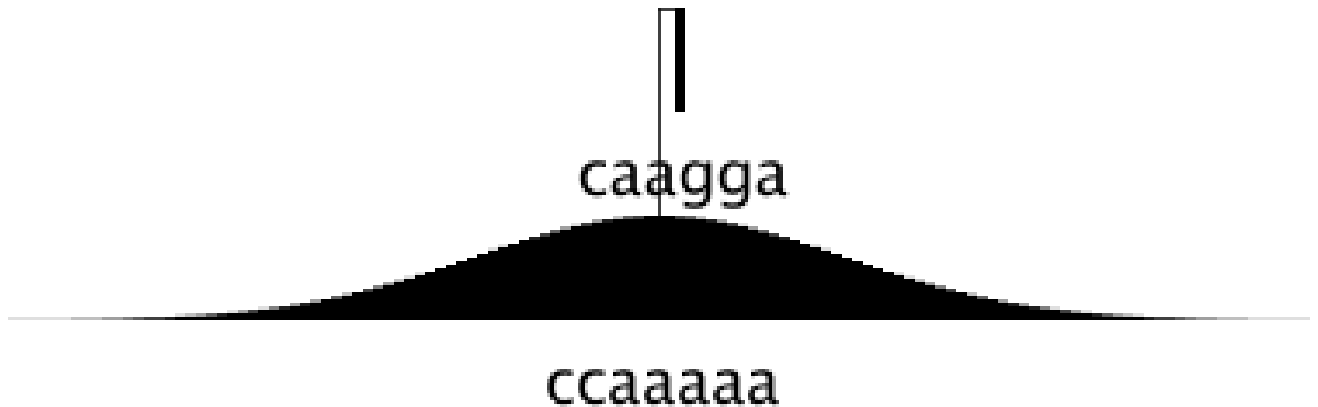}
\newline
\includegraphics[scale=0.2]{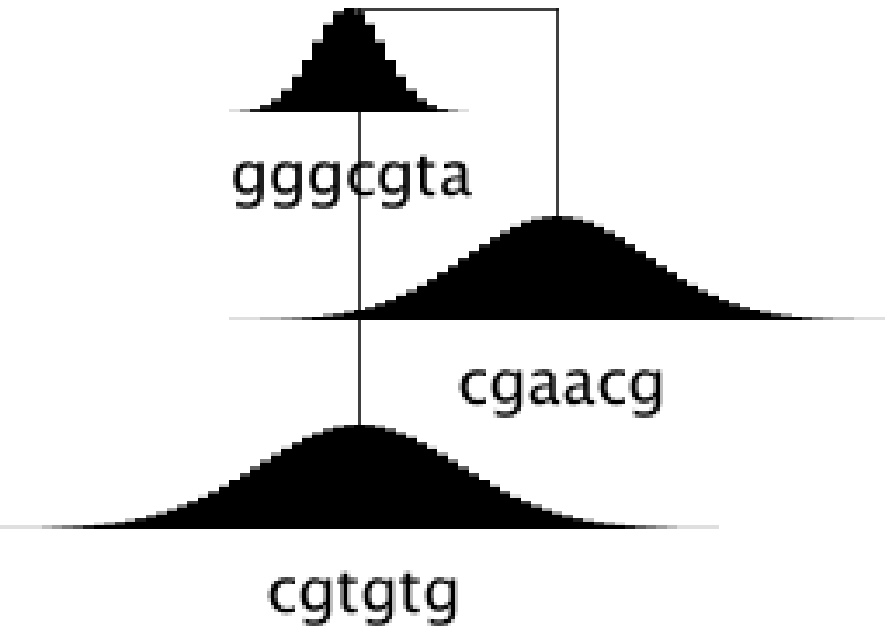}
\newline
\includegraphics[scale=0.2]{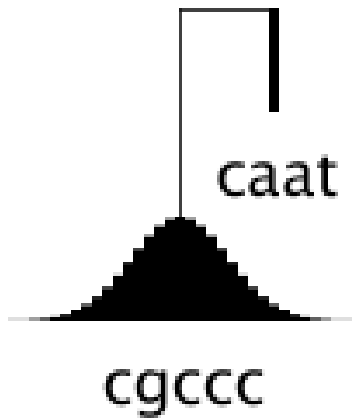}
\newline
\includegraphics[scale=0.2]{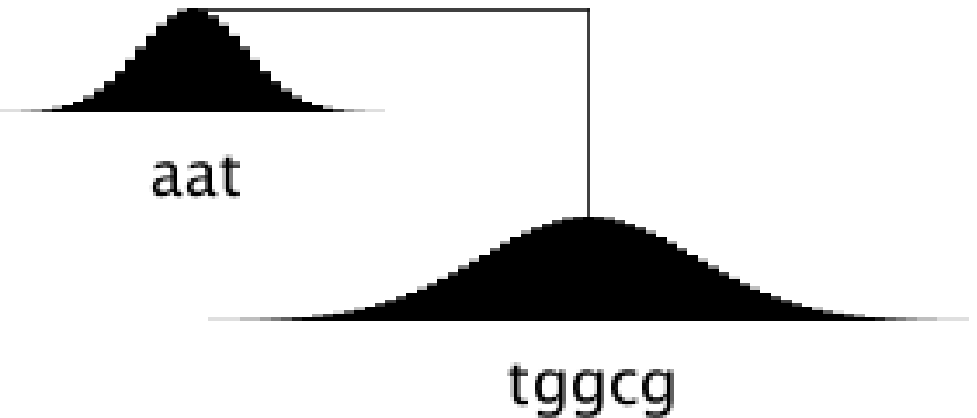}
\newline
\includegraphics[scale=0.2]{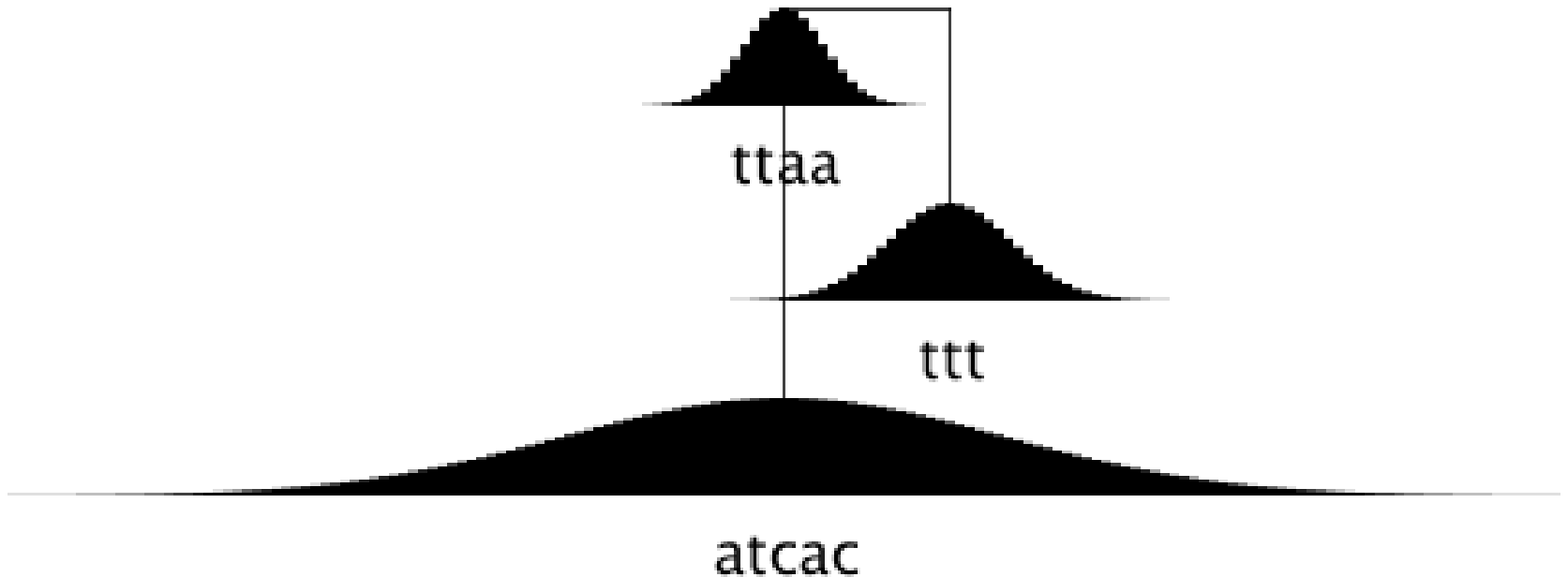}
\newline
\includegraphics[scale=0.2]{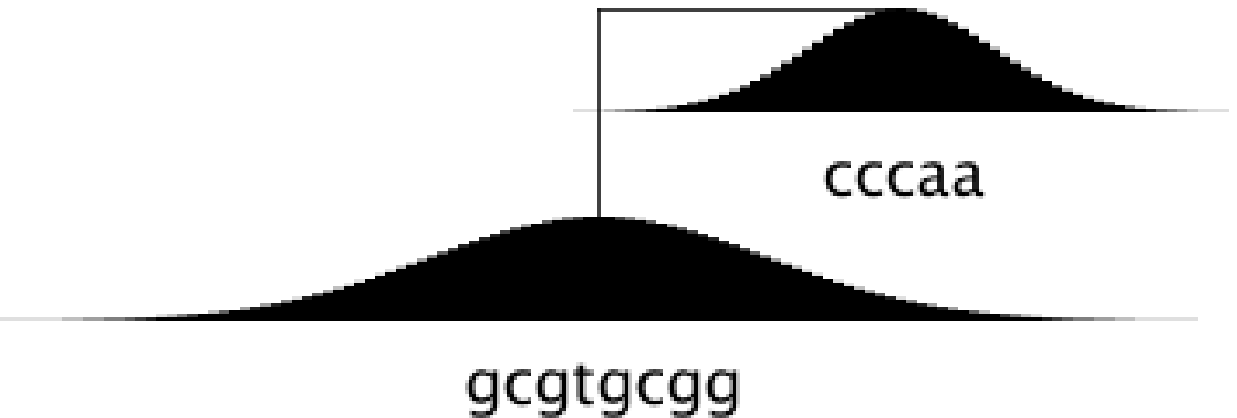}
\newline
\includegraphics[scale=0.2]{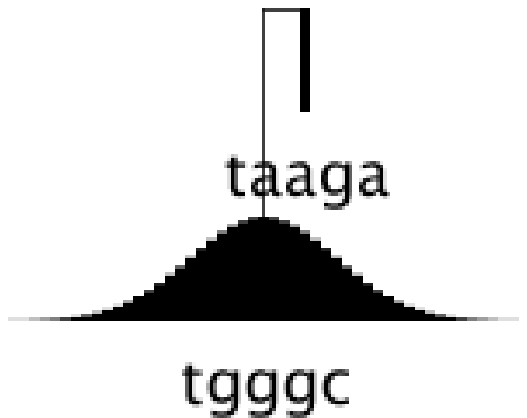}
\newline
\includegraphics[scale=0.2]{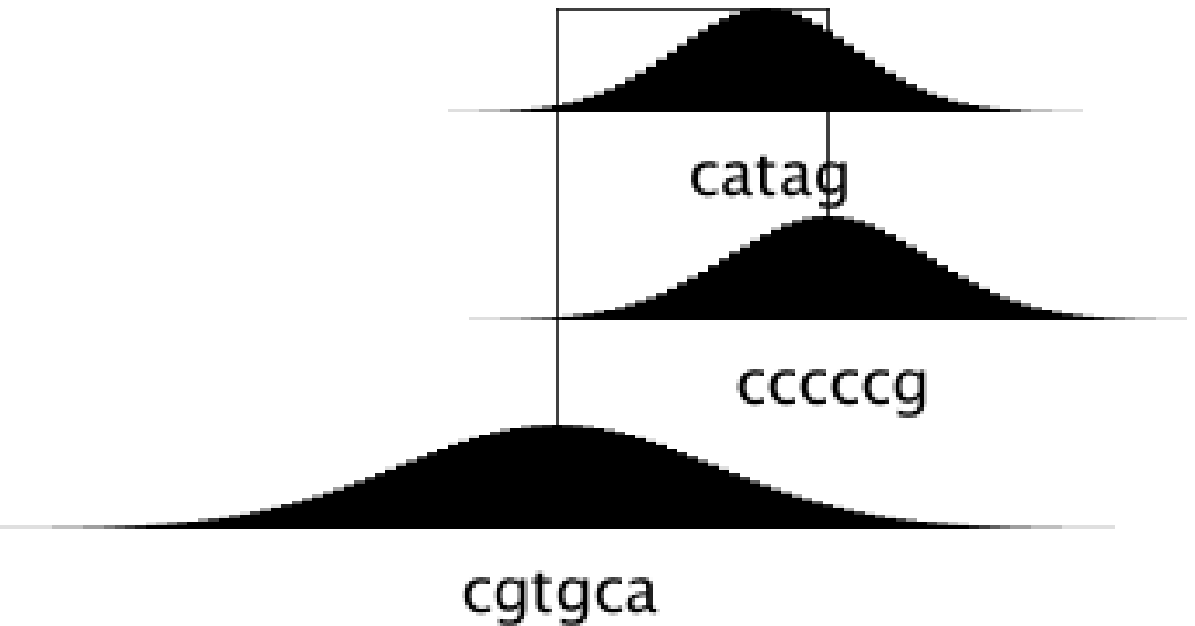}
\newline
\includegraphics[scale=0.2]{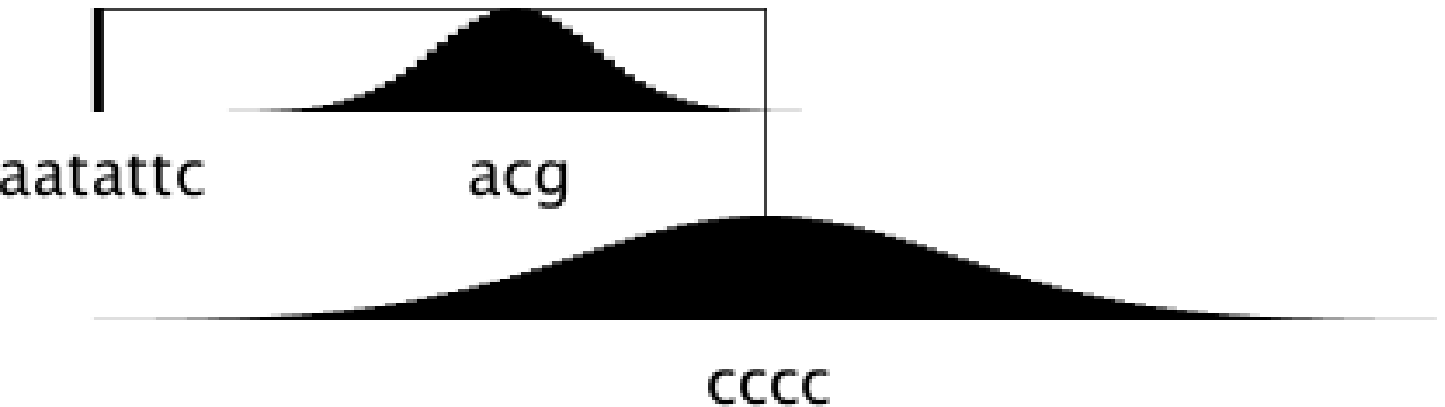}
\newline
\includegraphics[scale=0.2]{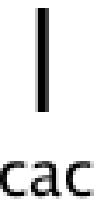}
\newline
\includegraphics[scale=0.2]{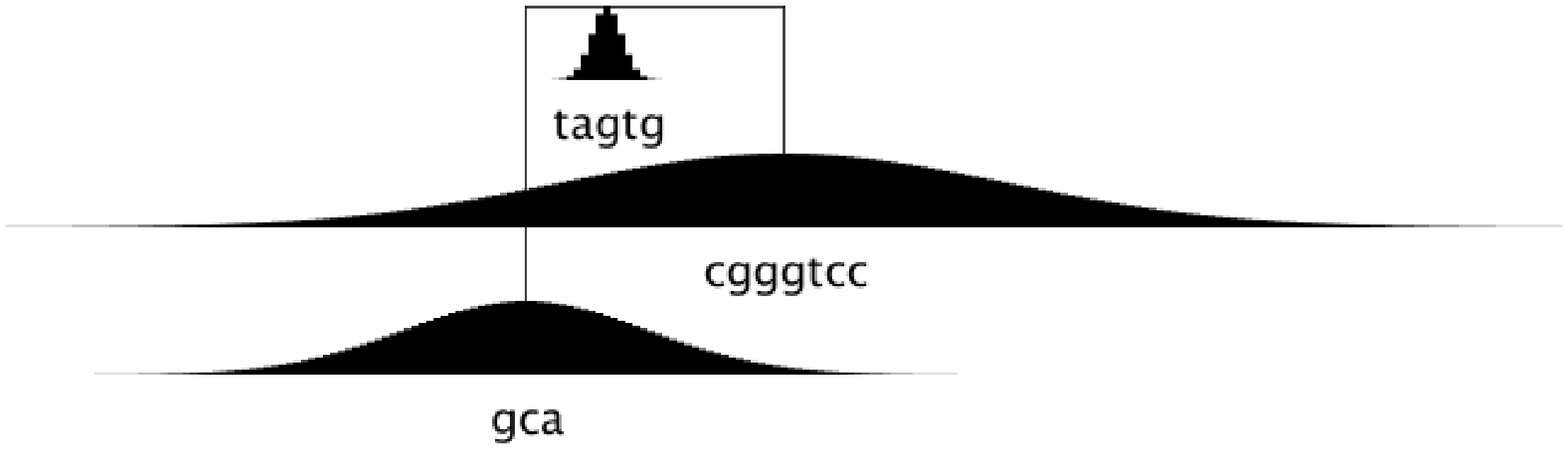}
\newline
\includegraphics[scale=0.2]{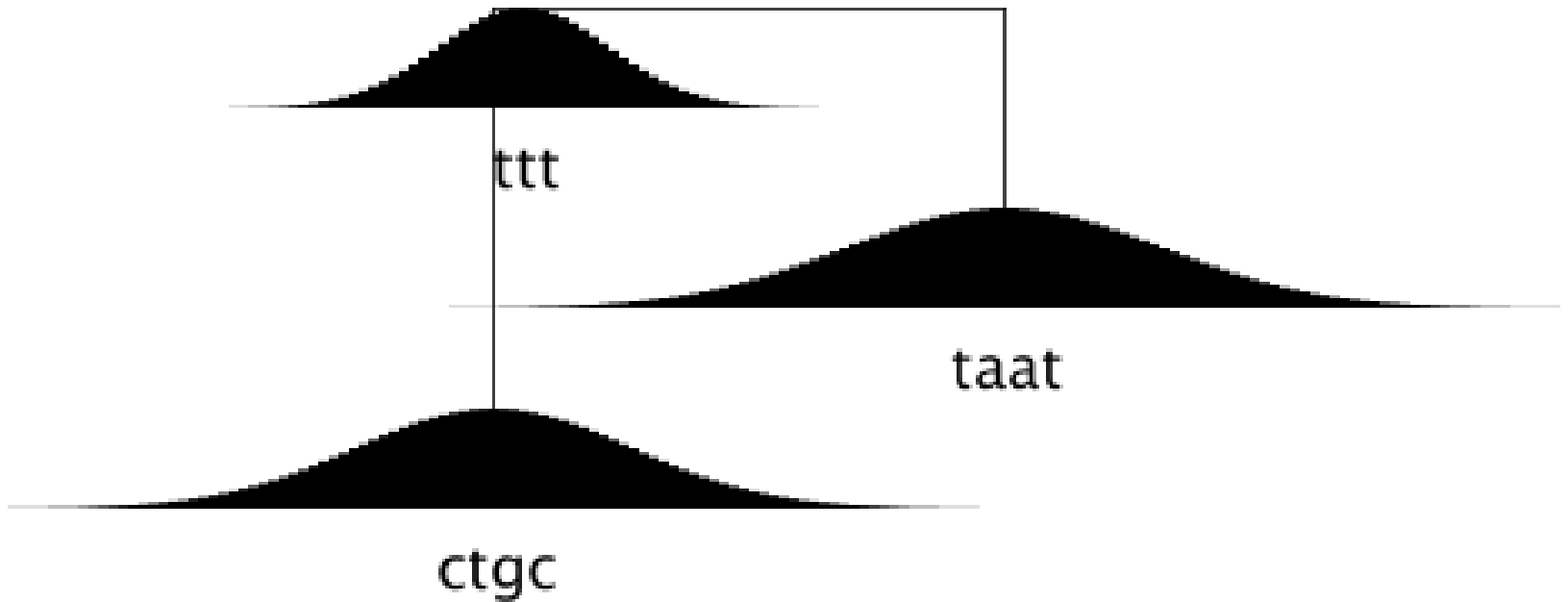}
&
\includegraphics[scale=0.2]{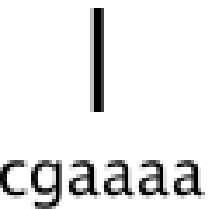}
\newline
\includegraphics[scale=0.2]{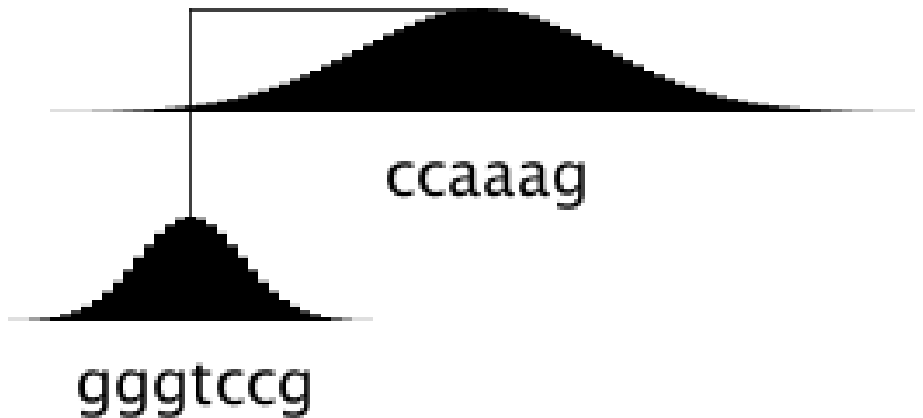}
\newline
\includegraphics[scale=0.2]{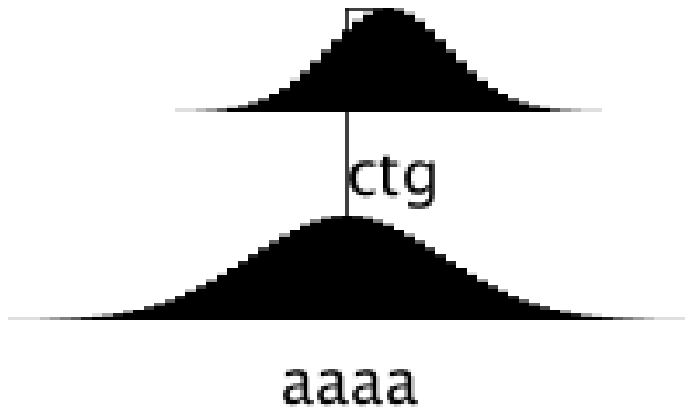}
\newline
\includegraphics[scale=0.2]{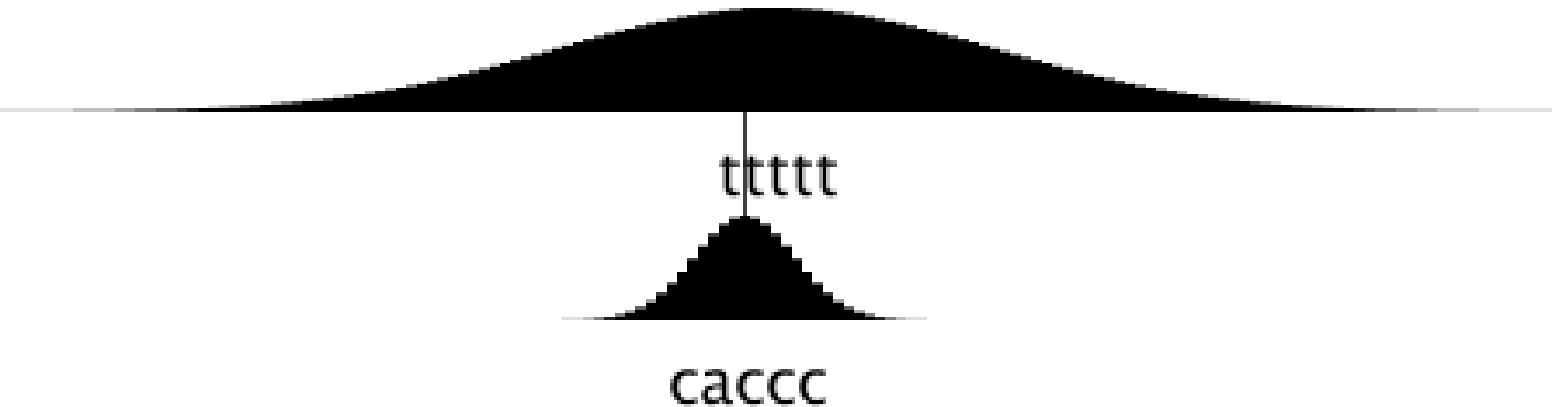}
\newline
\includegraphics[scale=0.2]{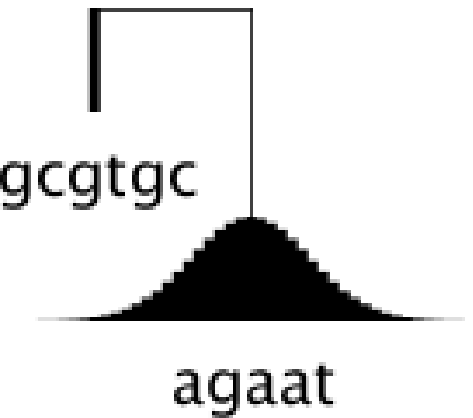}
\newline
\includegraphics[scale=0.2]{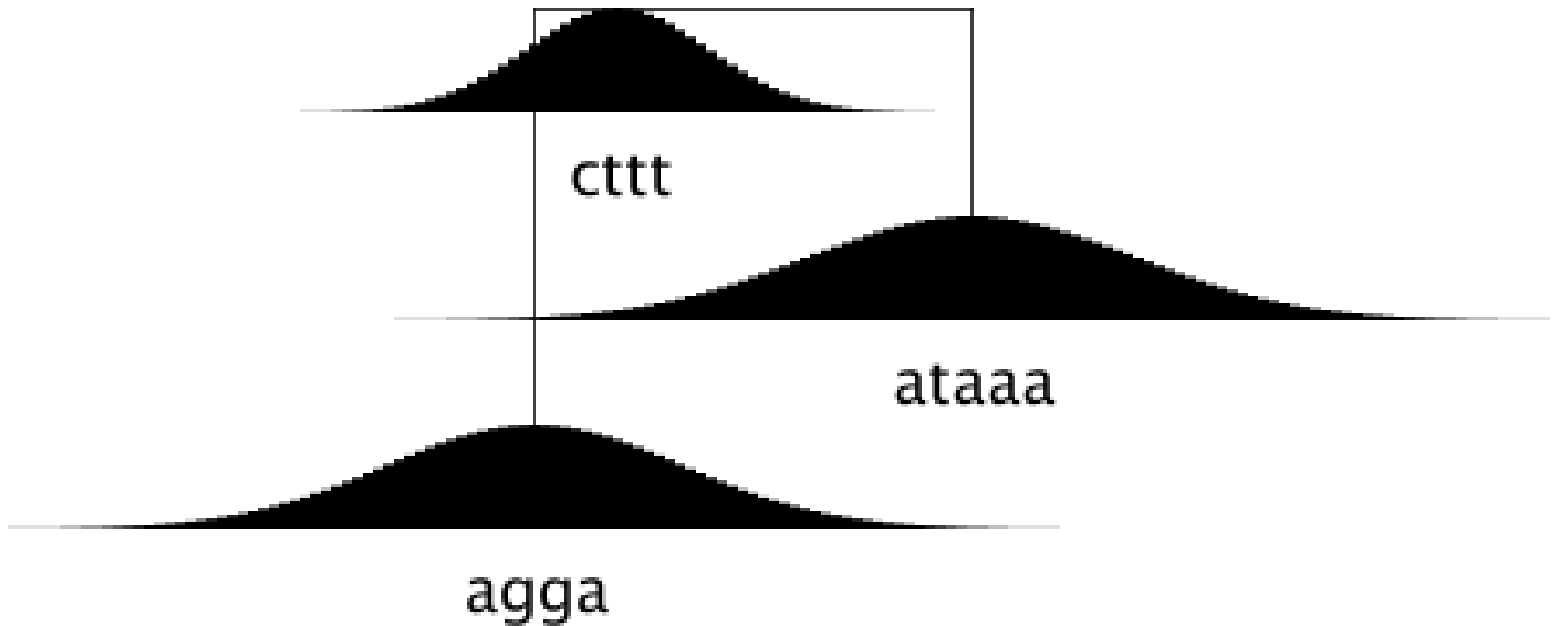}
\newline
\includegraphics[scale=0.2]{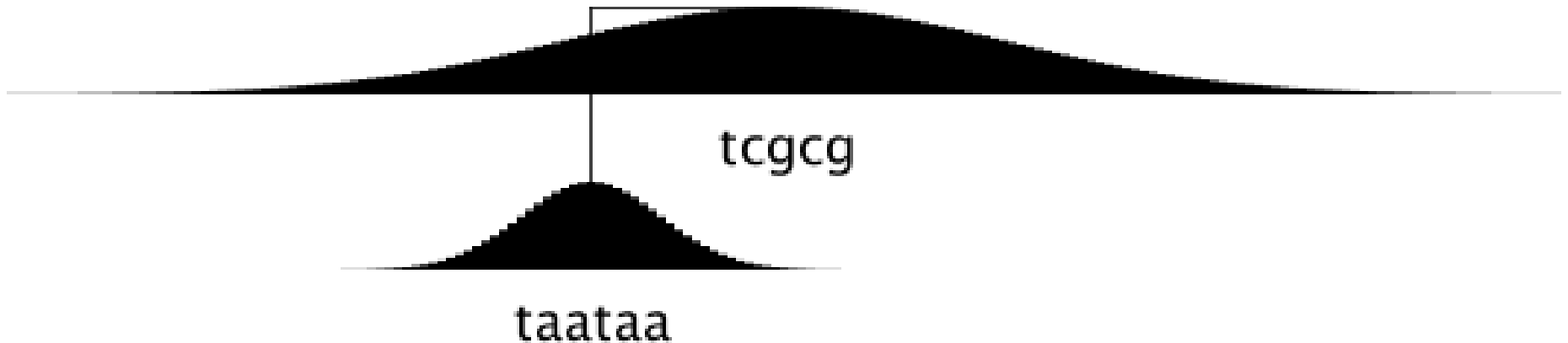}
\newline
\includegraphics[scale=0.2]{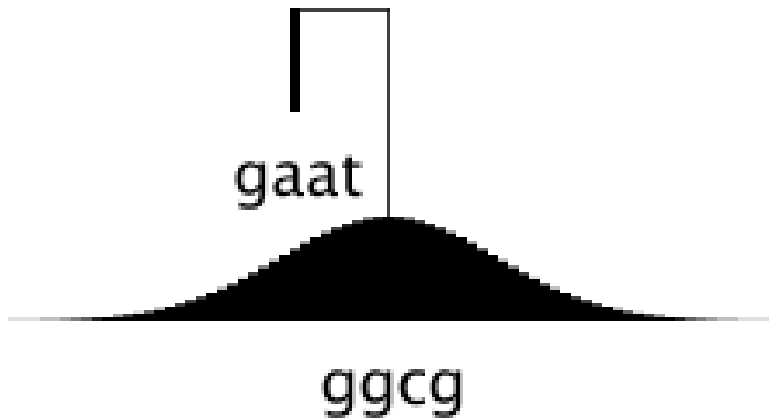}
\newline
\includegraphics[scale=0.2]{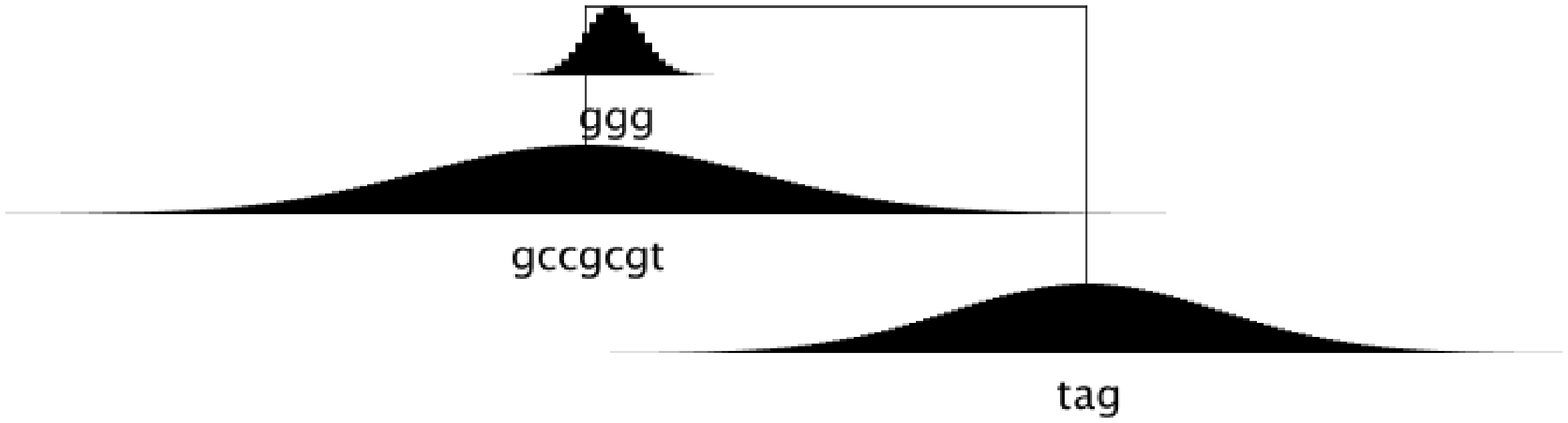}
\newline
\includegraphics[scale=0.2]{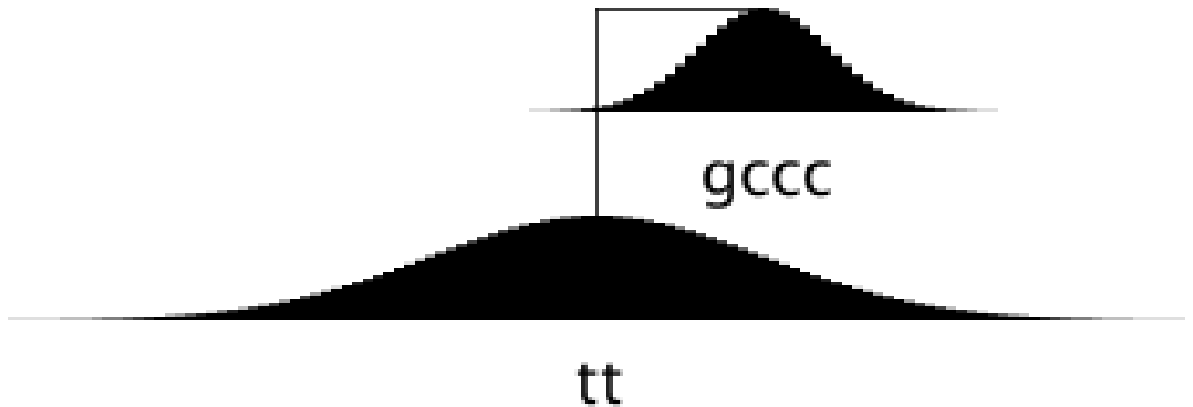}
\newline
\includegraphics[scale=0.2]{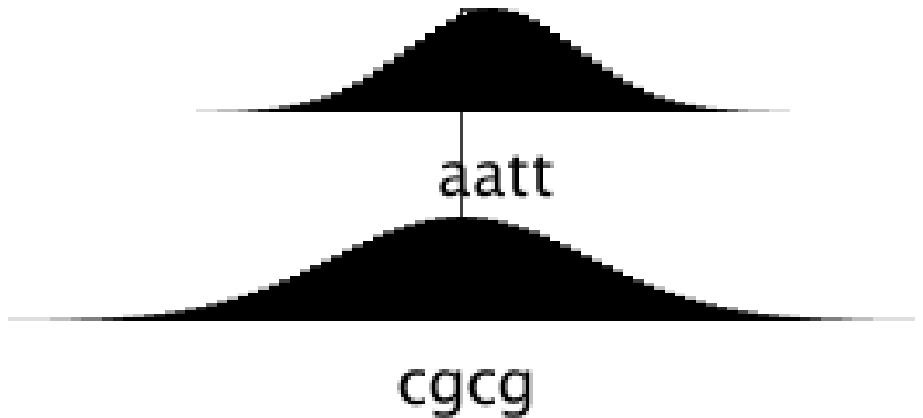}
\newline
\includegraphics[scale=0.2]{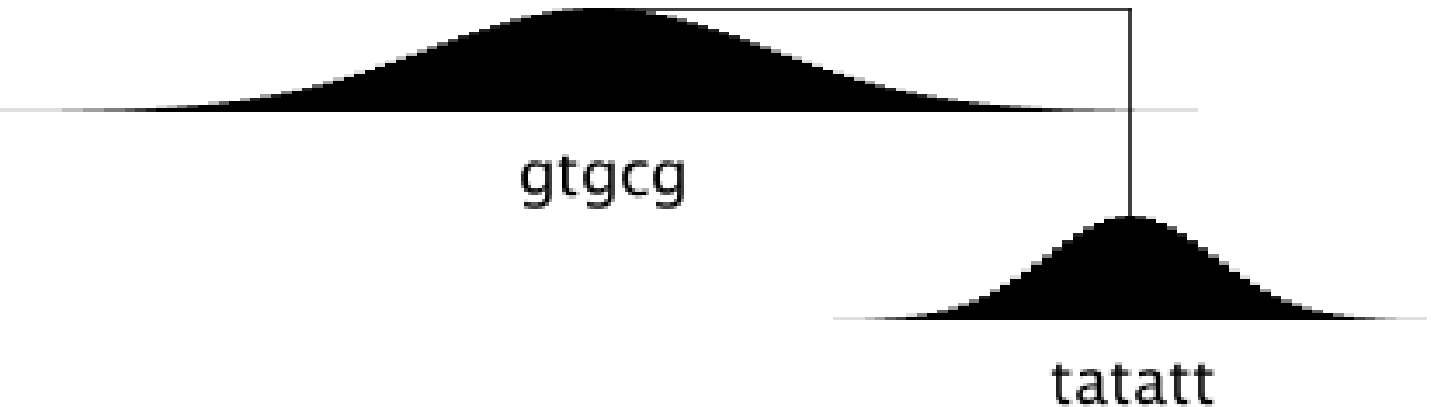}
\end{tabular}
\caption{Negatively weighted scaffolds in the Eponine Exons model}
\label{negative.scaffolds}
\end{center}
\end{figure}

This training procedure resulted in a complex model consisting of 216 scaffolds,
split evenly between positively-weighted scaffolds -- signals which are likely
to be over-represented in the true exons -- and negatively weighted scaffolds.  
The complete set of scaffolds can be seen in figures \ref{positive.scaffolds} 
and \ref{negative.scaffolds}.

\begin{figure}[!bth]
\begin{center}
\includegraphics[scale=1.0]{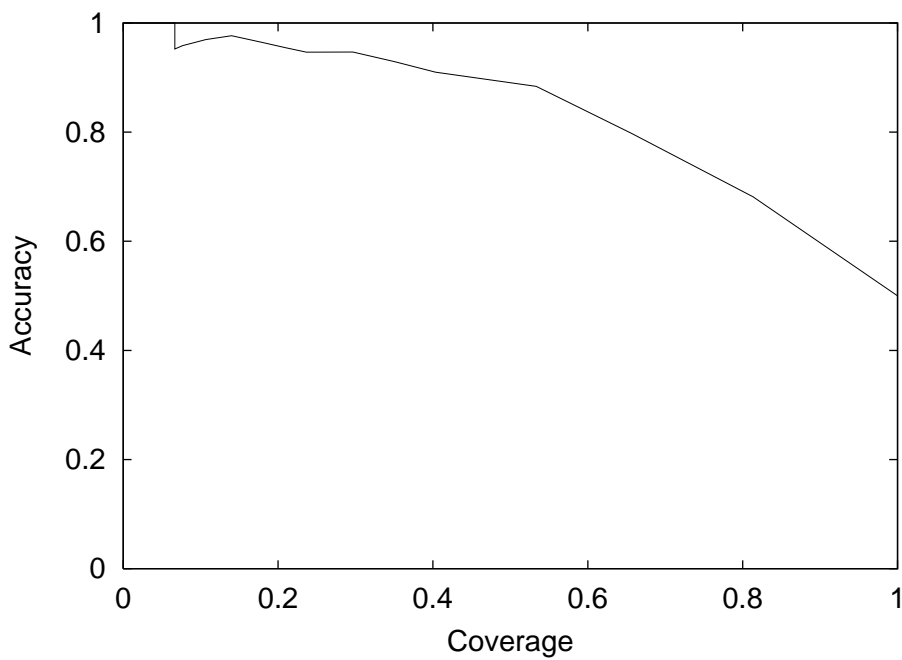}
\caption{Accuracy {\it vs.} coverage testing the model's ability to discriminate
  between unseen true and neutralized exons.}
\label{roc}
\end{center}
\end{figure}

We tested the resulting model's classification ability using the unseen data.
Accuracy (specificity or proportion of positive predictions which are correct) and coverage
(sensitivity or proportion of true exons detected) are shown for a range of classifier score
thresholds in figure \ref{roc}.  Clearly, the features learned by our procedure
are effective, in the general case, for distinguishing between true and neutralized
sequences.

\subsection{The eponine-exon model can also distinguish non-coding exons from randomized sequences}

Since we used semi-random sequences as the negative training set, an obvious
concern is that the features we have detected are artifacts of the neutralization
procedure, and are of no use when analysing real sequence data.
To validate the Eponine exon model, we tested it on additional sequences
from four classes: coding exons, non-coding (UTR) exons, introns, and intergenic
regions, all according to Vega annotation of finished human chromosomes.  In each case, we obtained a set of
1200 example sequences, each of 200 bases long.  For intergenic regions, we obtained
four independent sets of 1200 sequences.

\begin{figure}[!bth]
\begin{center}
\includegraphics[scale=1.0]{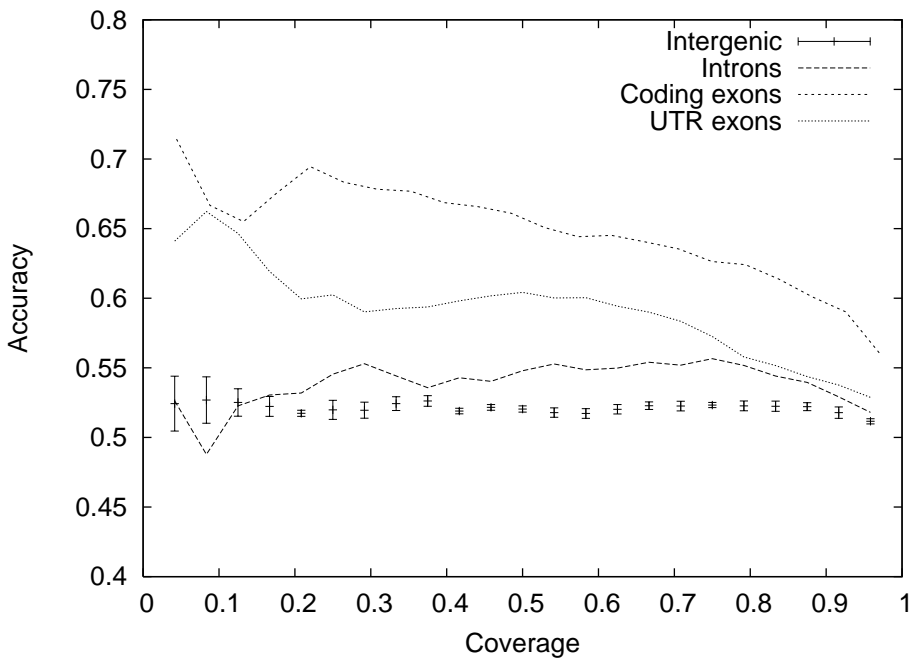}
\caption{ROC curves for the Eponine-Exons model on intergenic, intron, and UTR
exon sequences compared with random sequences of matching mono- and di-nucleotide
composition.  All curves are based on sets of 1200 sequences.  In the case of intergenic
sequences, standard-deviation error bars were calculated based on results from four
independent sets of sequences.}
\label{shuffle-rocs}
\end{center}
\end{figure}

For each data set, we produced a corresponding set of negative sequences with
matching mono- and di-nucleotide composition using the randomizing procedure
detailed in the methods section \ref{randomizing}.  We then used the Eponine-Exons model as
a classifier, and tested its ability to separate each of the positive sequence
sets from its corresponding negative sequence set.  Receiver Operating Characteristic
curves are shown in figure \ref{shuffle-rocs}.

In the case of the intergenic sequences, there is no significant discrimination
between real and shuffled sequences.  The coding sequences, however, could be
discriminated, as might be expected from a classifier trained on protein-coding
sequences.  However, the model was also able to distinguish many non-coding
exons from their shuffled counterparts.  This result is highly significant because
it indicates that at least some of the signals discovered in coding exons are actually
general to both coding and non-coding exons -- consistent with the idea that they
are involved in exon definition and splicing.  Finally, there is a far weaker,
but still possibility significant, discrimination of introns.  One explanation for
this is that the introns were contaminated with a small number of exons which
were missed during the annotation process.  However, a second possibility is
that, in addition to an exon-specific signal, the Eponine Exons model is also
detecting some (weak) signal -- perhaps an anti-termination signal -- which is found 
throughout transcribed regions of the genome.

\subsection{Comparison of learned motifs with known splice-enhancer sites}

We compared the weight matrices in the positively weighted scaffolds of our exon
model with known splice enhancer sites 
\citep{graveley,bourgeois.etal, liu.zhang.krainer,  lynch.etal, 
schaal.maniatis, tacke.manley, tacke.tohyama.ogawa, tian.kole, zheng.huynen.baker},
and also motifs detected by a very different computational approach, RESCUE-ESE
\citep{fairbrother.etal}.  Direct comparisons of weight matrices with sequence motifs --
with or without ambiguity symbols -- is complicated, since different positions in
a weight matrix may convey different amounts of information.  Furthermore, it
is not certain that either the learned weight matrices or the published motifs
correspond to the full length of the biologically functional sequence.  It it therefore
important to consider a range of possible alignments of motifs to weight matrices.

For each motif, we calculated the log-odds score against all weight matrices from
scaffolds with weights greater than 1.0, considering all possible alignments with
up to one base of overhang, and took the maximum score.  We then generated 500 shuffled
variants of the motif and scored these in the same fashion, taking the mean to be
a representative score for motifs of that particular length and base composition.
Tables \ref{motifexons}, \ref{motifintrons} and \ref{motifESE} lists the direct
and shuffled scores for experimentally determined motifs with our training
model, and also the difference between them. The status in the tables is to score a "+" or a "-" if 
the absolute observed difference is greater than $0.2$. Our model has managed to
predict some of the consensus sequences for exonic splicing enhancers that are
located within the internal exon (see table \ref{motifexons}) and did not detect any exonic splicing
enhancers consensus in the intronic regions near splice acceptor sites(see
table \ref{motifintrons}). The Eponine-Exons model successfully
distinguished the intronic and exonic consensus sequences for exonic
splicing enhancers.  Our model also finds some, but not all, of the motifs
detected by the RESCUE-ESE method.  This is consistent with the result above,
since RESCUE-ESE is designed to detect both exon-localized and intron-localized
motifs.

An unusual feature of our learning system is its ability to capture scaffolds of
related motifs, as well as individual motifs.  The scaffold (aagaatga
agcg ccccg) is particularly noteworthy. It is identified  
with the two known exonic splice enhancers, ASF/SF2 and SC35, which are known
to possess distinct, functionally significant RNA binding specificities \citep{tacke.manley}. 
In addition, the SR-related protein binding site for the Tra2$\beta$ is 
also recognized for the motif aagaatga. The mammalian Tra2 proteins are shown to
be sequence-specific activators of pre-mRNA splicing. The scaffold suggests that
there might be a connection between Tra2 proteins with the splice enhancers
ASF/SF2 and SC35. This might provide a starting point for predicting the
relationship between exonic splice enhancers and sequence specific activators.  

\begin{table}[!bth]
\begin{center}
\begin{tabular}{| p{2cm} | p{1.5cm} | p{1.2cm} |p{1.2 cm} |p{1.5 cm} |p{1.0 cm} | p{1.8 cm}  |}
\hline
Consensus & SR Protein  &    Direct Score  & Shuffled Score & Difference  & Status &
Reference \\ 
\hline
aggacagagc   & ASF/SF2   &  3.823  &  3.464  &  0.3589   &            
+ & Tacke et al (1995)   \\
aggacgaagc   & ASF/SF2   &  3.823 &  3.449  & 0.3737     &            
+ & Tacke et al (1995)  \\
rgaagaac    & ASF/SF2    &  3.207  &  3.296  &  -0.088 &				
  & Tacke et al (1995)   \\
acgcgca    &ASF/SF2    &  3.222   & 2.962 &  0.2602   &            
+ &	Tacke et al (1995)  \\
aggacrragc    &ASF/SF2    &  3.823  &  3.267  & 0.5564      &            
+  &	Graveley (2000) \\
tscgkm  &  SRp55  & 2.860 &  2.568 &   0.2915 &         
+	& Liu et al (1998)  \\
cctcgtcc    & SRp20   &  3.209  & 2.967 &  0.2420 &
+	&	Tacke et al (1999) \\
tgttcsagwt  &  SC35   &  3.484   & 2.877  & 0.6063  &
+	& Tacke et al (1999) \\
tgcngyy   &  SC35   &  2.748 &  2.622  & 0.1262	&
& Schaal et al (1999)  \\
acgaggay    & 9G8  &  3.508   & 3.137  &  0.3709          &        	
+   &	Graveley (2000)  \\
tcwwc   & dsx   &  2.553  & 2.146   & 0.4070           &       
+ &  Schaal et al (1999) \\ 
aggagat    &SC35   &  3.823   & 3.269   & 0.5546   &          
+ &	Graveley (2000)  \\
\hline
\end{tabular}
\caption{Comparison of known ESE motifs which are located inside the internal exons. 
The status in the tables is to score a "+" or a "-" if the absolute observed difference is greater than $0.2$.}
\label{motifexons}
\end{center}
\end{table}

\begin{table}[!bth]
\begin{center}
\begin{tabular}{| p{2cm} | p{1.5cm} | p{1.2cm} |p{1.2 cm} |p{1.5 cm} |p{1.0 cm}| p{1.8 cm}  |}
\hline
Consensus & SR Protein  &    Direct Score & Shuffled Score & Difference  & Status &
Reference   \\ \hline
ctcktcy   & SRp20   &        2.507 &  2.410 &  0.0976 &	
& Schaal et al (1999) \\
rgaccgg & SC35    &      3.064  & 3.065   &-0.001	&
& Schaal et al (1999)  \\
ggacaa    &ASF/SF2    &     2.507   & 3.096  &  -0.588       &           
- &	Schaal et al (1999)  \\
ggacag    &ASF/SF2     &     2.507   & 3.049  &  -0.541      &
- &	Schaal et al (1999) \\
agagcagg    &ASF/SF2     &    2.405   & 3.336  & -0.930       &          
- &	Zheng et al (1999) \\
rgackacgay    &9G8       &         2.196  &  3.035  & -0.838     &            
- & Tian et al (1999) \\
aagaagaa    &Tra2 (beta)  &      3.056  & 3.477  & -0.421          &       
- & Tacke et al (1995)  \\
tcaaca    &Tra2            &      2.287  & 2.726  & -0.439       &          
- &	Lynch et al (1996) \\
gaagaa    &hTra2 (beta)   &       3.056  & 3.393  & -0.337    &             
- & Graveley (2000)  \\
gacgacgag    &Pu1         &       2.296 &  3.347 &  -1.050    &             
- &	Bourgeois (1999) \\
gatgaagag   &Pu2         &       2.769 &  3.497  & -0.727     &           
- &	Bourgeois (1999) \\
\hline
\end{tabular}
\caption{Comparison of known ESE motifs which are located at the splice acceptor site of the exon-intron boundaries}
\label{motifintrons}
\end{center}
\end{table}

\begin{table}[!bth]
\begin{center}
\begin{tabular}{| p{1.8cm}  | p{1.2cm}  |  p{1.2 cm} |  p{1.5 cm} | p{1.1 cm}|}
\hline
Consensus &    Direct Score  & Shuffled Score & Difference  & Status  \\ \hline
atcttc  &      2.920 &  2.580  & 0.3400 &               
+  \\
actaca  &     1.982 &  2.729  &  -0.746   &              
- \\
ttggat   &      3.165  & 2.556   & 0.6090    &             
+ \\
gaatca &      3.328  &  3.040  &  0.2883   &              
+ \\
gaagaa  &     3.056  &  3.416 &  -0.360     &            
- \\
ttcaga    &     4.600   & 2.850  & 1.7501      &            
+  \\
gacaaa    &    2.922 &  3.093  &  -0.171     &     	
 \\
ctgaag    &    2.769  & 2.931   & -0.162	&			
 \\
aatcca    &   2.835  &  2.736  &  0.0984 &
    \\
aacttc    &    3.155  &  2.667  &  0.4877 &
+  \\
\hline
\end{tabular}
\caption{Comparison of known ESE motifs with the RESCUE-ESE method in
\cite{fairbrother.etal}}
\label{motifESE}
\end{center}
\end{table}

\section{Discussion}

We have shown that a motif-oriented machine learning strategy can extract signals
which discriminate effectively between true and neutralized sets of coding exons.
The resulting model included recognizable consensus sequences for many of the
previously reported splice-enhancer binding sites.  Although the model was trained
only on coding exon sequences, it gives high scores for both coding and non-coding
exons, but not introns or intergenic regions.  We therefore believe that the
neutralization strategy is a powerful and effective method for learning functional
non-coding elements embedded in protein coding sequence.

One interesting feature of the model learned here is its complexity: 216 scaffolds,
split evenly between positively and negatively-weighted scaffolds.  This is
a large number, both in absolute terms, and also in comparison with EWS and
C-EWS models trained for other purposes, such as promoter prediction (T. Down,
unpublished).  This suggests that a large number of functional elements play
widespread roles in exon definition.  Those motifs learned here which cannot
be assigned to any currently known splice-regulating protein are strong
candidates for investigation with a view to discovering novel splice regulators.
It may also be worth further investigation of the combination of motifs which
appear in scaffolds, since this could indicate interactions between proteins
in the splicing complex.

We hope that changes in the machine learning strategy will improve the classification
accuracy of this method.  Possible candidates for investigation include the use
of scaffolds comprising more than 3 motifs, and the replacement of simple
weight matrices with more complex models which serve as better representations
of protein binding sites.  We do not, however, necessarily expect that it will
be possible to classify true and neutralized exons with $100\%$ accuracy: most
proteins can accept many mutations with little or no change to structure and
function, so it is inevitable that some of the information which the cell uses
to define exons will be encoded in the choice of amino acids, rather that just
the choice of nucleotides used in redundant positions.

In the future, we hope to apply the results of this technique to the problem
of {\it ab initio} prediction of genes.  Current gene-prediction techniques
rely on a combination of splice-site models and `coding bias' -- the observation
that coding sequence looks very different from intronic and intergenic sequence
when considering properties such as hexamer frequencies.  While such methods
work reasonably well for protein-coding genes, they seldom make good predictions
of untranslated regions, and do not detect the non-coding RNA genes which
are now known to be important in many aspects of cellular function.  Scanning bulk
genomic DNA using our model makes many predictions outside known exons ({\it i.e.} a high
apparent false positive rate).  This suggests that while the motifs discovered here
may be necessary for efficient splicing, they are not sufficient to fully define
exons.  We hope that building knowledge of splice enhancers into gene prediction methods,
together with other features such as splice junction consensus sequence, will
improve the prediction of all spliced transcripts, whether coding or non-coding.

\section{Materials and methods} \label{methods}

\subsection{Genome sequence and annotation}

Human genome sequence release NCBI33  from Ensembl databases \citep{hubbard.ensembl}. 
Curated annotation of gene structures on chromosomes 6, 13, 14, 20,
and 22 were obtained from the Vertebrate Genome Annotation (Vega)
database [http://vega.sanger.ac.uk]. We extracted a total of 27954 internal translated
coding exons (see definition in \citealt{clark.thanaraj})
of different intron phases for our positive training set. Based on the definition 
\citep{clark.thanaraj}, an intron contained within CDS is said to have
a phase of zero if the intron demarcates a codon boundary, a phase of
one if it divides the codon between the first and second nucleotides,
and a phase of two if the intron divides a codon between the second and third
nucleotides. The position of an exon with respect to the codon
positions can be defined by the phases of upstream and downstream
flanking introns and when an exon is flanked by introns of the same
phase, it will be a multiple of three nucleotides in length. The phase
definition is important for the neutralization scheme described in
section \ref{neutralization}.
  
Vega and Ensembl data was extracted directly from the SQL databases using 
the BioJava toolkit with biojava-ensembl extensions [http://www.biojava.org/].

\subsection{Constructing a non-redundant set of sequences}

To eliminate similar sequences from the datasets, we performed an all-against-all
comparison of the sequences using NCBI blastn \citep{altschul.blast} using default options
(word size 11, match reward $+1$, mismatch penalty $-3$) and recorded all pairs
with a bit score $\geq 35$.  We then performed single-linkage clustering, and from
each cluster we picked one member at random to represent that cluster in the
final data set.

\subsection{Neutralization of coding sequences} \label{neutralization}

Exon neutralization is a process which randomizes the sequence of a set of
protein-coding exons while maintaining three key constraints:

\begin{itemize}
\item{The neutralized exons code for the same protein sequence as the real
exon}
\item{The frequency of a particular codon being used to represent a particular
amino acid is maintained}
\item{The overall dinucleotide composition of the set is maintained}
\end{itemize}

Thus, by comparing neutralized exons against the corresponding set of true
exons, it should be possible to detect larger sequence features and motifs
which are preferentially over- or under-represented in the true exon set.
Features which occur purely as artifacts of the underlying protein sequence,
or as a result of an overall preference to use particular codons, will occur
with equal frequency in the true and neutralized sets.

The neutralization process used here is a Monte-Carlo method, whereby small
(single-codon) changes to the sequence are proposed, then accepted or rejected
on the basis of a probabilistic model which captures the features listed above.
In this case, the model is encapsulated as a set of {\it conditional codon usage
tables}.  Consider a codon $C$ which encode amino acid $A$, and is flanked by
nucleotides $p$ and $q$ to form the pentanucleotide $pCq$.  Our model records
the probability of the codon being used in this context:

\begin{equation*}
P(C | A,p,q)
\end{equation*}

The model is initialized for a given set of exons by simply counting all in-frame
codons in the exon set.  Obviously, this means that a large data set is required
to construct a representative model, but the curated human gene set is sufficiently
large that this no longer presents a problem 

Now, for each exon in the set, a number of neutralization cycles are performed.
In each cycle, one in-frame codon position within the exon is chosen
at random. Let $C$ equal
the current codon at this position.  If it encodes
an amino acid which has only a single codon in the universal genetic code, it
is always left unchanged.  Otherwise, a synonymous codon, $C'$, is proposed by
sampling from a uniform random distribution over all synonyms, $Q(C' |
C)$.  Next, the appropriate  
conditional codon usage table is consulted, given the two bases either side
of $C$.  We accept or reject the proposed change on the basis of the 
Metropolis-Hastings criterion:

\begin{equation} \label{e:zeqn}
z = \frac{P(C')}{P(C)} \frac{Q(C' | C)} {Q(C | C')}
\end{equation}

When $z \geq 1$, the codon substitution is always accepted, when $z < 1$ the substitution is
accepted with probability $z$.  In this case, at any given position, the proposal
distribution $Q$ is always uniform, the second term of this expression can
be ignored: it is simply the fit of the proposed new codon to the model represented
by the conditional codon usage tables which is important. 

\subsection{Generating random sequences with matching mono- and di-nucleotide
  composition} \label{randomizing}
  
First, the sequence set is analysed and the initial dinucleotide composition
is recorded.  We then perform a large number (typically 500) iteration in which
two points within the sequence are selected at random, breaking it into three
segments, $ABC$.  We then propose a rearrangement to give the sequence
$BAC$.  This rearrangement destroys two dinucleotide pairings and creates
two new pairings.  The probabilities of the sequences $ABC$ and $BAC$ a calculated
from the dinucleotide frequency table, and the rearrangement is accepted or
rejected based on the Metropolis-Hastings criterion described above.

\subsection{The Eponine Windowed Sequence model} \label{eponine}

Convolved Eponine Window Sequence (C-EWS) models were trained using a Variational
Relevance Vector Machine as described in \citet{down.rvmseq}.  The model was seeded
using five-base motifs.  During the training process, sampling rules allowed
the motifs to be shortened, lengthened, or combined into scaffolds of up to
three motifs.

\section{Acknowledgments}

 We thank the Ensembl, HAVANA and the Chromosome 22 annotation group
for genome sequence and annotation data.
BL is supported by the International Fellowship from the Agency of
Science, Technology and Research (A-STAR), Singapore.
TD thanks the Wellcome Trust for support.

\end{document}